\documentclass[a4paper,usenames,dvipsnames,11pt]{article}
\usepackage{cite}
\usepackage[english]{babel}
\usepackage[makeroom]{cancel}
\usepackage{amssymb}
\usepackage{amsmath}
\usepackage{bbold}
\usepackage{booktabs, cellspace, hhline}
\usepackage{multirow}
\usepackage{graphicx}
\usepackage{hyperref}
\usepackage{dsfont}
\usepackage{xcolor}
\usepackage{setspace}
\usepackage{colortbl}
\newcommand{\Tr}{\textrm{Tr}}
\newcommand{\len}{\textrm{len}}

\usepackage{authblk}
\usepackage{mleftright}

\usepackage[mathscr]{euscript}
\newcommand{\sA}{\mathscr{A}}
\newcommand{\sB}{\mathscr{B}}
\newcommand{\sC}{\mathscr{C}}
\newcommand{\sD}{\mathscr{D}}
\newcommand{\sE}{\mathscr{E}}
\newcommand{\sF}{\mathscr{F}}
\newcommand{\sS}{\mathscr{S}}
\newcommand{\sP}{\mathscr{P}}
\newcommand{\sQ}{\mathscr{Q}}
\newcommand{\sR}{\mathscr{R}}
\newcommand{\sN}{\mathscr{N}}
\newcommand{\sM}{\mathscr{M}}

\newcommand{\sV}{\mathscr{V}}
\newcommand{\sW}{\mathscr{W}}
\newcommand{\sI}{\mathscr{I}}
\newcommand{\sJ}{\mathscr{J}}
\newcommand{\sAt}{\tilde{\mathscr{A}}}
\newcommand{\sBt}{\tilde{\mathscr{B}}}
\newcommand{\sCt}{\tilde{\mathscr{C}}}
\newcommand{\sDt}{\tilde{\mathscr{D}}}

\newcommand{\sFt}{\tilde{\mathscr{F}}}
\newcommand{\sSt}{\tilde{\mathscr{S}}}
\newcommand{\sPt}{\tilde{\mathscr{P}}}
\newcommand{\sQt}{\tilde{\mathscr{Q}}}
\newcommand{\sRt}{\tilde{\mathscr{R}}}
\newcommand{\sNt}{\tilde{\mathscr{N}}}
\newcommand{\sMt}{\tilde{\mathscr{M}}}

\newcommand{\sIt}{\tilde{\mathscr{I}}}
\newcommand{\sJt}{\tilde{\mathscr{J}}}

\newcommand{\U}{\text{U}}
\newcommand{\NC}{N_{\text{\tiny C}}}
\newcommand{\SU}{\text{SU}}
\newcommand{\M}{\mathcal{M}}
\newcommand{\A}{\mathcal{A}}

\begin{document}

\title{The colour matrix at next-to-leading-colour accuracy for tree-level multi-parton processes}
\author{Rikkert Frederix\footnote{rikkert.frederix@thep.lu.se}, Timea Vitos\footnote{timea.vitos@thep.lu.se}}

\affil[]{\small Theoretical Particle Physics, Department of Astronomy and Theoretical Physics, Lund University, S\"olvegatan 14A, SE-223 62 Lund, Sweden}

\date{}

\maketitle

\vspace*{-8.5cm}
{
     {LU-TP 21-42}
}
\vspace*{8.5cm}

\begin{center}
  \textbf{Abstract} \\
    \end{center}

\noindent  
We investigate the next-to-leading-colour (NLC)
contributions to the colour matrix in the fundamental and the
colour-flow decompositions for tree-level processes with all gluons,
one quark pair and two quark pairs. By analytical examination of the colour factors, we find the non-zero elements in the colour matrix at NLC. At this colour order, together with the symmetry of the
phase-space, it is reduced from factorial to polynomial the scaling of the contributing dual amplitudes as the number of partons participating in the scattering process is increased. This opens a path to an accurate tree-level matrix
element generator of which all factorial complexity is removed, without resulting to Monte Carlo sampling over colour.

\tableofcontents

\section{Introduction}
In the present time of preparation for the high-luminosity LHC runs, a
precise accounting for QCD background processes with high
multiplicities becomes increasingly important. Current tools for
calculating observables for high energy processes rely on perturbation
theory, and have been completely automated up to next-to-leading order
(NLO) accuracy in the QCD and EW coupling
constants~\cite{Alwall:2014hca,Frederix:2018nkq,Bothmann:2019yzt,Bellm:2015jjp,Alioli:2010xd}.

As the colour gauge group is one of the main bottlenecks in
matrix element calculations, there have been numerous progresses
towards structuring and simplifying the calculations. One of the first
main results was the recognition of colour decomposition of
amplitudes, in which the colour constants are stripped off the
amplitudes, leaving only the gauge-invariant dual amplitudes (also
called colour-ordered amplitudes) $\A_i$, dependent on the kinematics
only and all colour information is collected within a single matrix,
$C_{kl}$, in the colour-summed squared amplitude
\begin{eqnarray}\label{eq:colourdecomp}
|\M|^2 = \sum_{k,l} C_{kl} \A_k \A_l^*
\end{eqnarray}
where the sum runs over some generic subset of all permutations of the
dual amplitudes.  This method turns out however not to be unique: one
may choose a basis for the dual amplitudes in numerous ways,
some being minimal, some not. The most well-known bases are the
fundamental basis~\cite{Mangano:1987xk}, the colour-flow
basis~\cite{Maltoni:2002mq}, the adjoint basis~\cite{DelDuca:1999rs}
(all-gluon amplitudes only) and the recently introduced multiplet
basis~\cite{Keppeler:2012ih}.

When studying matrix elements for high-multiplicity parton scattering,
the number of dual amplitudes scales factorially with the number of partons involved in the interaction, and performing the double sum in
Eq.~\eqref{eq:colourdecomp} becomes the most time-consuming part of
the calculation, even longer than the actual evaluation of the dual
amplitudes. This problem can be mitigated by performing this sum using
Monte Carlo methods together with the phase-space sampling, as done for example in
Ref.~\cite{Gleisberg:2008fv}. This is typically done by assigning
randomly specific colours to the external partons involved in the
processes, which greatly reduces the number of dual amplitudes that
need to be considered per phase-space point. This method has a
drawback: it introduces larger point-by-point fluctuations in the Monte
Carlo integration of the phase-space, resulting in the need for more
phase-space points to be evaluated.

On the other hand, and somewhat less well-explored is the possibility
to perform a perturbative expansion in the number of colours,
$\NC$. In the large-$\NC$ limit~\cite{tHooft:1973alw} the gauge group
$\SU(\NC)$ is expanded to $\NC \rightarrow \infty$ while keeping
$g^2\NC$ fixed in the theory ($g$ being the coupling constant of the
gauge group). This method returns the same renormalisation group
equation for the corresponding coupling constant. It turns out that, apart from processes with
identical quarks, the odd powers in the expansion do not contribute,
resulting in an effective expansion parameter $1/\NC^2$. Its validity
within the Standard Model relies on its smallness with
$\NC=3$. Improving the expansion accuracy in this parameter increases
the precision in the large-$\NC$ limit, and the expanded terms are
dubbed leading-colour (LC), next-to-leading colour (NLC),
next-to-next-leading colour (NNLC), etc.

In this work, we investigate the large-$\NC$ expansion of the colour
matrix for tree-level multi-parton processes up to NLC accuracy. We will focus on all-gluon processes,
processes with a single quark line, and processes with two quark
lines\footnote{Processes with three or more quark lines are of less
importance from a phenomenological point of view. Even so, the results
presented in this paper can be extended to those cases as well.}. By
limiting to the NLC terms, the colour matrix $C_{kl}$ becomes sparse,
and, as we will show, the scaling of the complexity with the number of
partons involved in the process will reduce from factorial to
polynomial in the fundamental and colour-flow bases. It is mandatory
to be able to determine \emph{a priori} which terms in the colour
matrix contribute up to NLC accuracy, since computing the complete
colour matrix is undoable for high-multiplicity scattering processes.

The outline of this paper is the following. In
Sec.~\ref{sec:algorithms} and Sec.~\ref{sec:CF} we present the NLC
expansion of the colour matrix in the fundamental and colour-flow
decompositions, respectively. In Sec.~\ref{sec:results} we present our
results in terms of the number of dual (conjugate) amplitudes that
need be computed for all-gluon processes, processes with one quark
pair, and processes with two quark pairs. In Sec.~\ref{sec:conclusion}
we summarise and discuss the results.

\section{The colour matrix in the fundamental decomposition}\label{sec:algorithms}

In order to proceed with the presentation of the next-to-leading
colour terms, we summarise a few of the very useful identities related
to the fundamental generators of $\SU(\NC)$. We pick the normalisation
of the fundamental $T^a$ matrices to be
\begin{eqnarray}
\Tr[T^a T^b] = T_R\delta_{ab}.
\end{eqnarray}
with the index $T_R=1$ for convenience. The Fierz identity
\begin{eqnarray}\label{eq:Fierz}
(T^a)_{ij} (T^{a})_{kl} = \delta_{il}\delta_{jk}-\frac{1}{\NC}\delta_{ij}\delta_{kl}
\end{eqnarray}
incorporates the tracelessness and Hermiticity of the matrices and is key to any
simplifications of colour factors. We will use Einstein summation convention whenever a colour index is explicitly written out. For the fundamental
matrices, the identity leads to the two well-known relations
\begin{align}\label{eq:rule1}
\textrm{Rule~I:}&\qquad\qquad\Tr[T^a \sR ]\Tr[T^a \sS ] = \Tr[\sR\sS]- \frac{1}{\NC} \Tr[\sR] \Tr[\sS] , \\\label{eq:rule2}
\textrm{Rule~II:}&\qquad\qquad\Tr[\sR T^a \sQ T^a \sS ] = \Tr[\sQ] \Tr[\sR\sS] - \frac{1}{\NC} \Tr[\sR\sQ\sS],
\end{align}
where $\sR$, $\sS$ and $\sQ$ are strings of generators of arbitrary
length or the identity matrix $\mathbb{1}$. We shall denote the length
of a string $\sS$, i.e., the number of generators in the string, by
$\len(\sS)$, and if the string is the identity matrix, then
$\len(\sS)=0$. Our proofs about the (N)LC contributions, that we will
present below, rely on the application of these relations multiple
times such that we end up with several products of multiple
$\Tr[\mathbb{1}]=\NC$. Note that Rule~I reduces two traces into a
single trace (first term), and therefore reduces the possibility to
generate many $\Tr[\mathbb{1}]=\NC$, or leaves the number of traces
the same (second term), but with an additional factor of
$1/\NC$. Hence, this rule effectively reduces the potential maximum
power of $\NC$ in the final result. Rule~II increases the number of
traces (first term), keeping the potential power of $\NC$ unchanged,
or keeps the number of traces the same, but with an additional $1/\NC$
factor in front. Hence, only the first term of Rule~II does not
decrease the potential maximum power of $\NC$'s. This implies that the
most-leading contribution in the colour expansion comes from strings
(in which all colour indices are summed over) that have a form such
that it can be computed using only
\begin{align}\label{eq:rule2b}
  \textrm{Rule~IIb:}\qquad\qquad\Tr[\sR T^a T^a \sS ] = \NC
  \Tr[\sR\sS] + \mathcal{O}(1/\NC)
\end{align}
repeatedly. Hence, for any string of length $\len(\sP)=2p$ with $p$ different gluon colour indices, we have
$\Tr[\sP]=\NC^{p+1}+\mathcal{O}(\NC^{p-1})$ if it can be simplified
using Rule~IIb alone.

We also define the rules of Eqs.~\eqref{eq:rule1}~and~\eqref{eq:rule2} for the string
versions,
\begin{align}\label{eq:rules_long}
\begin{split}
  \Tr[\sA \sR] \Tr[\sAt \sS] & =
  \frac{1}{\NC}\left[\left(\NC-\frac{1}{\NC}\right)^a-\left(\frac{-1}{\NC}\right)^a\right] \Tr[\sR\sS] +
  \left(\frac{-1}{\NC}\right)^{a} \Tr[\sR]\Tr[\sS],\\
  \Tr[\sR \sA \sQ \sAt \sS] & =
  \frac{1}{\NC}\left[\left(\NC-\frac{1}{\NC}\right)^a-\left(\frac{-1}{\NC}\right)^a\right] \Tr[\sQ]\Tr[\sR\sS] +
  \left(\frac{-1}{\NC}\right)^{a} \Tr[\sR\sQ\sS]
\end{split}
\end{align}
where the sum is implied over all generator indices in the string
$\sA$ with $\len(\sA)=a$ not containing any repeated indices and $\sA$
is not the identity matrix, while $\sQ$, $\sR$ or $\sS$ can be the
identity matrix. We denote the mirrored string with a tilde, so $\sAt$
has the reverse order for gluon indices to $\sA$. Throughout this
work, we use the convention that if a repeated string appears, all the
colour indices of the fundamental generators are summed over in that
string.

\subsection{$n$-gluon amplitude}\label{sec:gluon}
The colour-summed squared matrix element, Eq.~\eqref{eq:colourdecomp}, for all-gluon
processes in the fundamental basis is given by
\begin{eqnarray}\label{eq:squared_gluons}
|\M|^2 =  (g^2)^{n-2}\sum_{\sigma_k,\sigma_l \in S_{n-1}} C(\sigma_k,\sigma_l) \A (\sigma_k(1),\ldots ,\sigma_k(n-1),n)\left(\A(\sigma_l(1),\ldots ,\sigma_l(n-1),n)\right)^*
\end{eqnarray}
with the colour matrix
\begin{eqnarray}\label{eq:matrix_gluons}
C(\sigma_k,\sigma_l) = \Tr[ T^{a_{\sigma_k(1)}} \ldots  T^{a_{\sigma_k(n-1)}} T^{a_n} ]  \left(\Tr[T^{a_{\sigma_l(1)}} \ldots  T^{a_{\sigma_l(n-1)}}  T^{a_n}  ]\right)^*
\end{eqnarray}
with an implicit sum over all gluon indices. This is an $(n-1)! \times
(n-1)!$ matrix. The expansion of these colour factors in powers of $\NC$
is the focus of the discussions below.

The dual amplitudes in the fundamental (and colour-flow decompositions)
are not independent and form an over-complete basis. The amplitudes
are related by a set of relations (the Kleiss-Kuijf relations~\cite{Kleiss:1988ne}), such as the dual Ward
identity at tree level
\begin{eqnarray}\label{eq:DWI}
\A(1,2,3,4\ldots ,n) + \A(2,1,3,4,\ldots ,n) + \ldots  + \A(2,3,4,\ldots ,1,n ) = 0,
\end{eqnarray}
from which it can be shown that the number of independent, gauge-invariant dual amplitudes reduces to $(n-2)!$. Such a minimal basis is used in the adjoint decomposition~\cite{DelDuca:1999rs} for all-gluon amplitudes. \footnote{Work on developing minimal bases for processes with quark lines has been pursued in Refs.~\cite{Reuschle:2013qna,Johansson:2015oia,Melia:2015ika,Ochirov:2019mtf}.}
\subsubsection*{Leading colour}
The leading-colour contribution is obtained when the two permutations in Eq.~\eqref{eq:squared_gluons} are the same, $\sigma_k=\sigma_l$. Letting the string of generators in
the permutation be $\sN$ of length $n$, the colour factor is
\begin{eqnarray}\label{eq:fund_gluon_LC}
  \Tr[\sN](\Tr[\sN])^* =
 \Tr[\sN]\Tr[\sNt] =
  \left(\NC-\frac{1}{\NC}\right)^n+(\NC^2-1)\left(\frac{-1}{\NC}\right)^n =
  \NC^n+\mathcal{O}(\NC^{n-2}).
\end{eqnarray}

As introduced earlier, owing to the dual Ward identity in Eq.~\eqref{eq:DWI}, not all of
the dual amplitudes in this expansion are independent. Using the
relation, one may re-write the full-colour expansion in terms of a
subset of the dual amplitudes and for this reason, for $n=4,5$ gluons, the
leading-colour contributions in the colour matrix, with a suitable
polynomial in $\NC$, recover the full-colour accurate
result~\cite{Mangano:1987xk}. In the present work, we do not consider
the dual Ward identity to further simplify the number of terms
needed to obtain NLC accurate result, although it is a topic to
consider for further developments.

\subsubsection*{Next-to-leading colour}
The next-to-leading-colour (NLC) contribution, i.e.~of order
$\mathcal{O}(\NC^{n-2})$, is obtained if the two strings of
fundamental generators in the traces of the colour factor appearing in Eq.~\eqref{eq:matrix_gluons} may
be written as
\begin{eqnarray}\label{eq:NLC_gluon_Fund_gen}
\Tr[\sR\sQ_1 \sS \sQ_2 \sP]   \left(\Tr[\sR \sQ_2 \sS \sQ_1 \sP]\right)^* 
\end{eqnarray}
for the amplitude and conjugate amplitude respectively. \footnote{This was previously worked out in the approximation of a pure $\U(\NC)$ gluon in Ref.~\cite{Bern:1990ux}.} That is, the
permutations of the amplitude and its conjugate are the same up to
interchanging a single string of generators. In other terms, the permutations are related by block interchanges~\cite{Christie}. Here, $\sR$, $\sS$ and $\sP$
are strings of generators of arbitrary length (including zero), and
$\len(\sQ_{1,2})\ge 1$.  By cyclicity and the form of the colour factor also
$\len(\sR) \ge 1$ holds. Explicit computation of the colour factor
yields that the value is always equal to
$+\NC^{n-2}+\mathcal{O}(\NC^{n-4})$ up to a few exceptions. The
conditions and/or exceptions to this rule are the following, all for the case in which
$\sQ_1$ and $\sQ_2$ are neighbouring strings, i.e.~$\sS=\mathbb{1}$ ($\len(\sS)=0$).
\begin{itemize}
\item If $\len(\sQ_1)=\len(\sQ_2)=1$ the colour factor is a NLC term equal
  to $-\NC^{n-2}+\mathcal{O}(\NC^{n-4})$.
\item If $\len(\sQ_1)=1$ \emph{or} $\len(\sQ_2)=1$ the colour factor is a
  NLC term equal to $-\NC^{n-2}+\mathcal{O}(\NC^{n-4})$ if $\len(\sR)
  = 1$ and $\len(\sP) = 0$. If the conditions on $\sR$ and $\sP$ are
  not met, this is not a NLC contribution.
\item If $\len(\sQ_{1,2})>1$ the colour factor is a NLC term equal to
  $+\NC^{n-2}+\mathcal{O}(\NC^{n-4})$ if $\len(\sR) \neq 1$ and
  $\len(\sP) \neq 0$. If the conditions on $\sR$ and $\sP$ are not
  met, this is not a NLC contribution.
\end{itemize}
Finally, as can be seen from Eq.~\eqref{eq:fund_gluon_LC}, that if the
permutations in the dual amplitude and the complex conjugate are
identical, i.e.~a LC term, the colour factor also contributes at NLC
with $-n\NC^{n-2}$.

\subsubsection*{Proof of the NLC terms}\label{sec:fund_gluon_proof}
Let us consider the interference of an amplitude with permutation
$\sigma_k$ and string of generators $\sN_k$ and a conjugate amplitude
with permutation $\sigma_l$ and string of generators $\sN_l$, both
strings being of size $\len(\sN_{k,l})= n$. The colour factor from Eq.~\eqref{eq:matrix_gluons} is
\begin{eqnarray}
C(\sigma_k,\sigma_l)  = \Tr[\sN_k]\left(\Tr[\sN_l]\right)^*=\Tr[\sN_k]\Tr[\sNt_l].
\end{eqnarray}
Without loss of generality, we may use the cyclicity of the trace to
rearrange the above strings to a form in which at least one generator
is common in the most-left position of the string, and possibly some
common generators at the end,
\begin{eqnarray}
  \sN_k=\sR\sM_k\sP, \quad\quad \sN_l=\sR\sM_l\sP.
\end{eqnarray}
The lengths of the strings are $\len(\sR)=r\ge 1$, $\len(\sP)=p\ge 0$ and
let $\len(\sM_k)=\len(\sM_l)=m$. The LC and NLC contributions are
then of order $\mathcal{O}(\NC^{r+p+m})$ and
$\mathcal{O}(\NC^{r+p+m-2})$, respectively. After using the relations
from Eq.~\eqref{eq:rules_long} to sum over all generator indices in
$\sR$ and $\sP$ we rewrite the colour factor as
\begin{multline}\label{eq:full}
\Tr[\sR\sM_k\sP]\Tr[\sPt\sMt_l\sRt] = \frac{1}{\NC}\left[
  \left(\NC-\frac{1}{\NC}\right)^{r+p}
  -\left(\frac{-1}{\NC}\right)^{r+p} \right]\Tr[\sM_k\sMt_l] +
\left(\frac{-1}{\NC}\right)^{r+p} \Tr[\sM_k]\Tr[\sMt_l].
\end{multline}
Only the first term can contribute at LC accuracy, and only if the
$\Tr[\sM_k\sMt_l]$ can be completely reduced using Rule~IIb
(Eq.~\eqref{eq:rule2b}). The final term can contribute to NLC only
in the case of $r=1$ and $p=0$.

The cases $\sM_{k,l}=\mathbb{1}$ and $\sM_{k,l}=T^a$ are trivial cases
that lead to LC colour factors.  Hence we only need to consider
$\len(\sM_{k,l}) \ge 2$. Let the first and final gluon indices for the
generators in the string $\sM_k$ be $a$ and $b$, i.e.,
\begin{equation}
\sM_k=T^a \sE T^b.
\end{equation}
We distinguish the following two cases for $\sM_l$:
\begin{align}\label{eq:cases}
\sM_{l}=  \begin{cases} 
\sA T^a \sB T^b \sC   \quad \text{with } \quad \sA,\sC \neq \mathbb{1} \quad \quad &\text{(case 1)}\\
\sD T^b \sS T^a \sF &\text{(case 2)}.
\end{cases} 
\end{align}
Note that for case~1 the string $\sA$ ($\sC$) cannot be equal to
$\mathbb{1}$, since otherwise the $T^a$ ($T^b$) would be part of $\sR$
($\sP$).

We will start by considering the first case. Within this first case,
we first limit ourselves to the situation in which $r>1$ and/or
$p>0$. Therefore, the only possible NLC contribution is the first term
in Eq.~\eqref{eq:full} which, upon using Rules~I~\&~II
(Eqs.~\eqref{eq:rule1}~and~\eqref{eq:rule2}) simplifies to
\begin{equation}\label{eq:case1}
\Tr[\sM_k\sMt_l]=\Tr[\sAt]\Tr[\sCt]\Tr[\sE\sBt]-\frac{1}{\NC}\Tr[\sAt]\Tr[\sE\sCt\sBt]- 
\frac{1}{\NC}\Tr[\sCt]\Tr[\sE\sBt\sAt]+\left(\frac{1}{\NC}\right)^2\Tr[\sE\sCt\sBt\sAt].
\end{equation}
All of these terms can only contribute at NNLC or beyond. The reason
for this is that the $\len(\sE)=m-2$. Hence, with $\sE$ appearing in a
trace, only the repeated application of Rule~IIb
(Eq.~\eqref{eq:rule2b}) to that trace can give at most
$\mathcal{O}(\NC^{m-1})$, which would exactly be a NLC
contribution. Since $\sA\neq \mathbb{1}$ and $\sC \neq \mathbb{1}$,
the first three terms of Eq.~\eqref{eq:case1} cannot be simplified by
that rule alone, since at some point the generators in $\Tr[\sAt]$
and/or $\Tr[\sCt]$ need to be combined with the generators in
$\Tr[\sE]$ using Rule~I (Eq.~\eqref{eq:rule1}). Potentially, only the
trace in the final term of Eq.~\eqref{eq:case1} could give a
$\mathcal{O}(\NC^{m-1})$ colour factor if $\sE=\sA\sB\sC$. However,
that term comes with a $1/\NC^2$ factor in front resulting in a NNLC or higher order term.

In the situation in which $r=1$ and $p=0$, we must investigate the
additional second term in Eq.~\eqref{eq:full} which gives
\begin{multline}
-\frac{1}{\NC} \Tr[\sM_k]\Tr[\sMt_l]=\\-\frac{1}{\NC}\left(\Tr[\sE\sBt]\Tr[\sAt\sCt]-\frac{1}{\NC} \Tr[\sE\sAt\sCt\sBt]- \frac{1}{\NC}\Tr[\sE\sBt\sAt\sCt]+\left(\frac{1}{\NC}\right)^2 \Tr[\sE] \Tr[\sBt\sAt\sCt]\right).
\end{multline}
Using the same argument as above, none of these terms can give NLC
contributions. Hence, we have ruled out the possibility for case~1 in
Eq.~\eqref{eq:cases} to yield a NLC contribution.

We move now on to case~2 in Eq.~\eqref{eq:cases}.  We explicitly evaluate
Eq.~\eqref{eq:full} using Rules~I~\&~II
(Eqs.~\eqref{eq:rule1}~and~\eqref{eq:rule2}), arriving at
\begin{multline}\label{eq:gluon_trace}
\frac{1}{\NC}\left[
  \left(\NC-\frac{1}{\NC}\right)^{r+p}
  -\left(\frac{-1}{\NC}\right)^{r+p} \right]\\ \times \Bigg[\Tr[\sFt\sE\sDt\sSt]-\frac{1}{\NC}\Tr[\sFt\sE]\Tr[\sDt\sSt]-
\frac{1}{\NC}\Tr[\sFt\sSt]\Tr[\sDt\sE]+\cancel{\left(\frac{1}{\NC}\right)^2 \Tr[\sE\sFt\sSt\sDt] }\Bigg] \\
+\left(\frac{-1}{\NC}\right)^{r+p} \Bigg[\Tr[\sSt]\Tr[\sDt\sFt\sE]+ \cancel{\left(\frac{-1}{\NC}\right) \Tr[\sE\sSt\sDt\sFt]} + 
\cancel{\left(\frac{-1}{\NC}\right)\Tr[\sSt\sDt\sFt\sE]}+\cancel{\left(\frac{1}{\NC}\right)^2\Tr[\sE]\Tr[\sDt\sFt\sSt]}\Bigg].
\end{multline}
The crossed out terms cannot possibly contribute to NLC due to the
already contracted two indices in $\sM_{k,l}$ and the several $1/\NC$
prefactors. For the remaining terms, in order for them to contribute at
NLC, it must be possible to simplify them using Rule~IIb
(Eq.~\eqref{eq:rule2b}) alone. For the first term of the middle line,
this is the case if and only if $\sE=\sF\sS\sD$; for the second term
if and only if $\sE=\sF$ and $\sD=\sS=\mathbb{1}$; for the third term
if and only if $\sE=\sD$ and $\sF=\sS=\mathbb{1}$; and for the fourth
term (i.e., the first term in the lower line) if and only if $r+p=1$
and $\sE=\sD\sF$ and $\sS=\mathbb{1}$. Hence, the first term gives the
general form of the NLC terms, Eq.~\eqref{eq:NLC_gluon_Fund_gen}, with
$\sQ_1=T^a\sF$ and $\sQ_2=\sD T^b$. For the exceptions in the case
$\sQ_1$ and $\sQ_2$ are neighbouring strings, i.e.~corresponding to
$\sS=\mathbb{1}$, the second to fourth terms also contribute, and due
to the relative minus sign difference between these terms and the
first, this can either lead to no NLC contribution, or a negative NLC
contribution\footnote{It cannot lead to a $-2$ NLC contribution, since
for that all four terms must contribute. This is not possible since
for this $\sE=\sF=\sS=\sD=\mathbb{1}$ and $\len(\sR)+\len(\sP)=1$, and
this corresponds to a 3-gluon scattering amplitude.}. This completes
the proof.

\subsection{One $q\overline{q}$ pair and $n$ gluons}\label{sec:fund_1qq}

The colour-summed, squared amplitude with one quark pair and $n$ gluons
is given by \cite{Mangano:1987kp,Mangano:1988kk,Mangano:1990by,Dixon:1996wi,Caravaglios:1998yr}
\begin{eqnarray}\label{eq:squared_one}
|\M_{1qq}|^2 =  (g^2)^{n}\sum_{\sigma_k,\sigma_l \in S_n}C(\sigma_k,\sigma_l)  \A_{1qq}(q,\overline{q},\sigma_k(1),\ldots ,\sigma_k(n))\left(\A_{1qq}(q,\overline{q},\sigma_l(1),\ldots ,\sigma_l(n))\right)^*
\end{eqnarray}
with the colour matrix 
\begin{eqnarray}\label{eq:colorone}
C(\sigma_k,\sigma_l)= \Tr[T^{a_{\sigma_k(1)}}\ldots
  T^{a_{{\sigma_k(n)}}}T^{a_{\sigma_l(n)}}\ldots T^{a_{\sigma_l(1)}}].
\end{eqnarray}

\subsubsection*{Leading colour}
The leading-colour contributions are those for which the permutations in Eq.~\eqref{eq:squared_one} are the same,
$\sigma_k=\sigma_l$. Denoting the string of fundamental generators in the order of the permutation as $\sN$ (with $\len(\sN)=n$), the colour factor for this LC contribution is
\begin{eqnarray}\label{eq:fund_1qq_LC}
\Tr [\sN\sNt] =
  \NC \left(\NC -\frac{1}{\NC} \right)^n =
  \NC^{n+1}+\mathcal{O}(\NC^{n-1}).
\end{eqnarray}

\subsubsection*{Next-to-leading colour}
Similarly to the all-gluon case, the colour factors in Eq.~\eqref{eq:colorone} will yield NLC of
$\mathcal{O}(\NC^{n-1})$ contribution if and only if the two strings
of generators for the two permutations can be written as
\begin{eqnarray}\label{eq:1qq_Fund_gen}
  \Tr[ \sR \sQ_1 \sS \sQ_2 \sP \sPt \sQt_1 \sSt \sQt_2 \sRt ].
\end{eqnarray}
That is, the permutations of the amplitude and its conjugate are the
same up to interchanging two strings of generators. Here, $\sR$,
$\sS$ and $\sP$ are strings of generators of arbitrary length (including
zero), and $\len(\sQ_{1,2})\ge 1$. These permutations yield a NLC
contribution equal to $+\NC^{n-1}+\mathcal{O}(\NC^{n-3})$ with the following
exceptions.
\begin{itemize}
\item If $\sS=\mathbb{1}$ and $\len(\sQ_1)=\len(\sQ_2)=1$ the colour factor is a NLC term equal to $-\NC^{n-1}+\mathcal{O}(\NC^{n-3})$.
\item If $\sS=\mathbb{1}$ and $\len(\sQ_1)=1$ \emph{or}
  $\len(\sQ_2)=1$ the colour factor is not a NLC term, but
  $\mathcal{O}(\NC^{n-3})$.
\end{itemize}
Furthermore, as can be seen from Eq.~\eqref{eq:fund_1qq_LC}, if
the permutations in the dual amplitude and the complex conjugate are
identical, i.e.~a LC term, the colour factor also contributes at NLC with
$-n\NC^{n-1}$.

\subsubsection*{Proof of the NLC terms}
The proof for the one-quark-line amplitude follows very closely to
that for the all-gluon amplitude presented in
Sec.~\ref{sec:fund_gluon_proof}. The colour factor is now a single
trace of two strings of generators corresponding to the two
permutations as given in Eq.~\eqref{eq:colorone},
\begin{eqnarray}
  \Tr[\sN_k \sNt_l].
\end{eqnarray}
Once again, we may simplify this by stripping off the beginning and
end of strings, which possibly coincide. Note that we now cannot
use cyclicity to always assure that $\len(\sR)\ge 1$,
\begin{eqnarray}
  \sN_k = \sR \sM_k \sP \quad,\quad \sN_l = \sR\sM_l \sP,
\end{eqnarray}
which yields the simplified colour factor
\begin{eqnarray}
  \Tr[\sR\sM_k \sP \sPt\sMt_l\sRt] = \left(\NC-\frac{1}{\NC}\right)^{r+p} \Tr[\sM_k \sMt_l].
\end{eqnarray}
The colour factor is LC if $\sM_k = \sM_l$ with the order
$\mathcal{O}(\NC^{r+p+m+1})$. This form is very similar to that in
Eq.~\eqref{eq:full} with only the first term present. Hence, we expect
the same set of rules to apply, but without the exceptions which
appeared when adding the second term for $r=1$ and $p=0$ in the
all-gluon case. The procedure is to again express $\sM_{k,l}$ in
the two most general forms following the steps in
Sec.~\ref{sec:fund_gluon_proof}. We exclude case~1 in
Eq.~\eqref{eq:cases} by the same arguments as presented in that
section, leaving us only with case~2:
\begin{eqnarray}
  \sM_k = T^a \sE T^b,  \quad\quad \sM_l = \sD T^b \sS T^a \sF.
\end{eqnarray}
Contracting the two indices $a$ and $b$ yields precisely the
expression~\eqref{eq:gluon_trace} with only the middle line,
\begin{eqnarray}\label{eq:gluon_trace_four}
  \Tr[\sFt\sE\sDt\sSt]-\frac{1}{\NC}\Tr[\sFt\sE]\Tr[\sDt\sSt]-
  \frac{1}{\NC}\Tr[\sFt\sSt]\Tr[\sDt\sE]+\cancel{\left(\frac{1}{\NC}\right)^2 \Tr[\sE\sFt\sSt\sDt] }
\end{eqnarray}
to be considered. As before, these traces yield NLC if and only if
they can be simplified using Rule~IIb alone. Hence, the first term
gives the general form of the NLC terms, Eq.~\eqref{eq:1qq_Fund_gen},
while the second and third terms lead to the following exceptions.
\begin{itemize}
\item $\sD=\sS=\sF=\mathbb{1}$: the colour factor is $-\NC$, which
  contributes at the NLC.
\item $\sD=\sS=\mathbb{1}$ and $\sF\neq \mathbb{1}$: all possible NLC terms
  cancel.
\item $\sF=\sS=\mathbb{1}$ and $\sD\neq \mathbb{1}$: all possible NLC terms
  cancel.
\end{itemize}
This completes the proof.

\subsection{Two distinct-flavour $q\overline{q}$ pairs and $n$ gluons}\label{sec:two_DF}

We first consider the case with two distinct flavour quark
lines. Further discussions on the multiple quark case can be found in,
e.g.~Refs.~\cite{Kosower:1988kh,Bern:1994fz,Ellis:2008qc,Ellis:2011cr,Ita:2011ar,Melia:2013bta,Melia:2013epa,Melia:2014oza}. In general for multi-quark amplitudes,
the dual amplitudes are no longer the partial amplitudes, meaning that
each colour factor multiplies not a single dual amplitude, but rather
a sum of them. In the case of two quark lines, however, the partial amplitudes coincide with the dual amplitudes and therefore one may speak again of dual amplitudes only without referring to the partial amplitudes. The notation typically presented in the literature for
multiple-quark line amplitudes, see e.g.~Ref.~\cite{Schuster:2013aya},
is not very well suited for the same-flavour case, that will be
discussed in Sec~\ref{sec:fund_2qq_SF}.  In this work, we introduce a
notation that adapts to both the distinct-flavour and same-flavour
amplitudes.

The matrix element for the two distinct-flavour quark lines is
decomposed as per usual
\begin{eqnarray}
\M_{2qq} = \sum_{\sigma} c(\sigma) \A_{2qq}(\sigma)
\end{eqnarray}
with a set of dual amplitudes $\A_{2qq}(\sigma)$ and $\sigma$ is some
permutation over a set of the gluon and quark indices, which we
specify below. Looking only at the (anti-)quark orderings, this
amplitude can be decomposed in the following two colour-ordered
diagrams (not placing out any of the external gluons explicitly)
\begin{equation}\label{eq:2qq_decomp}
\vcenter{\hbox{\includegraphics[scale=0.3]{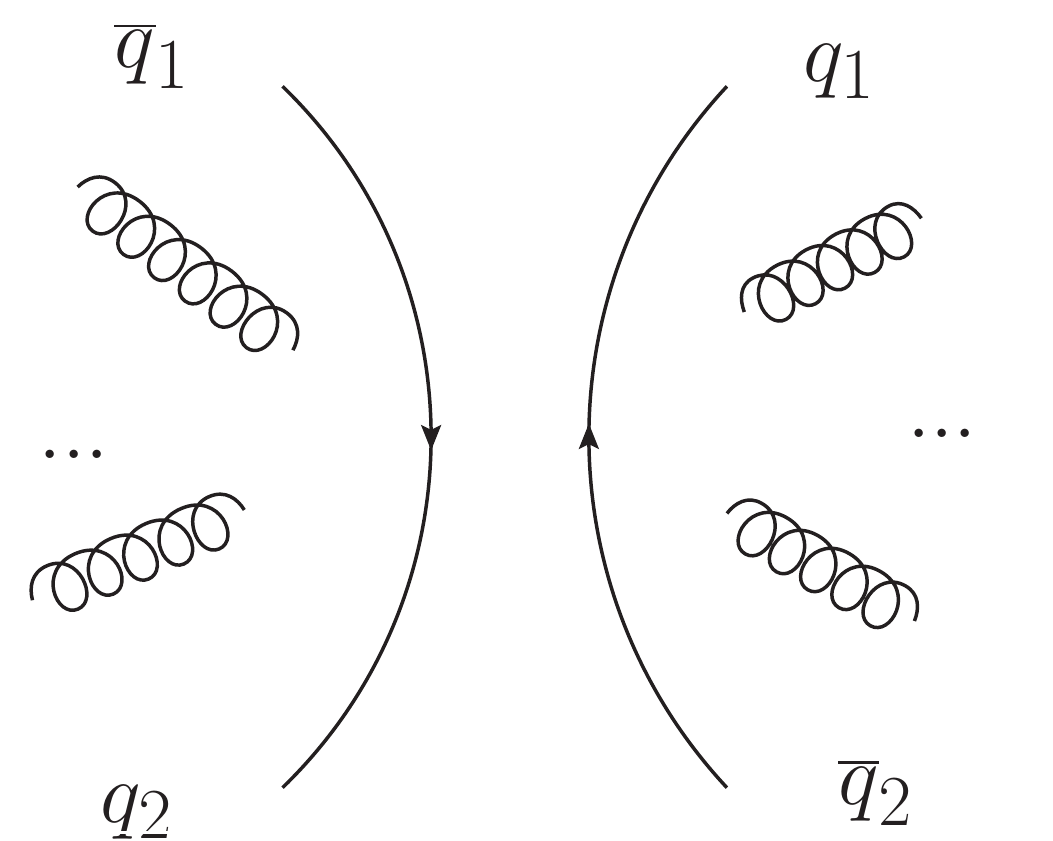}}} \quad-\,
\frac{1}{\NC}\,\vcenter{\hbox{\includegraphics[scale=0.3]{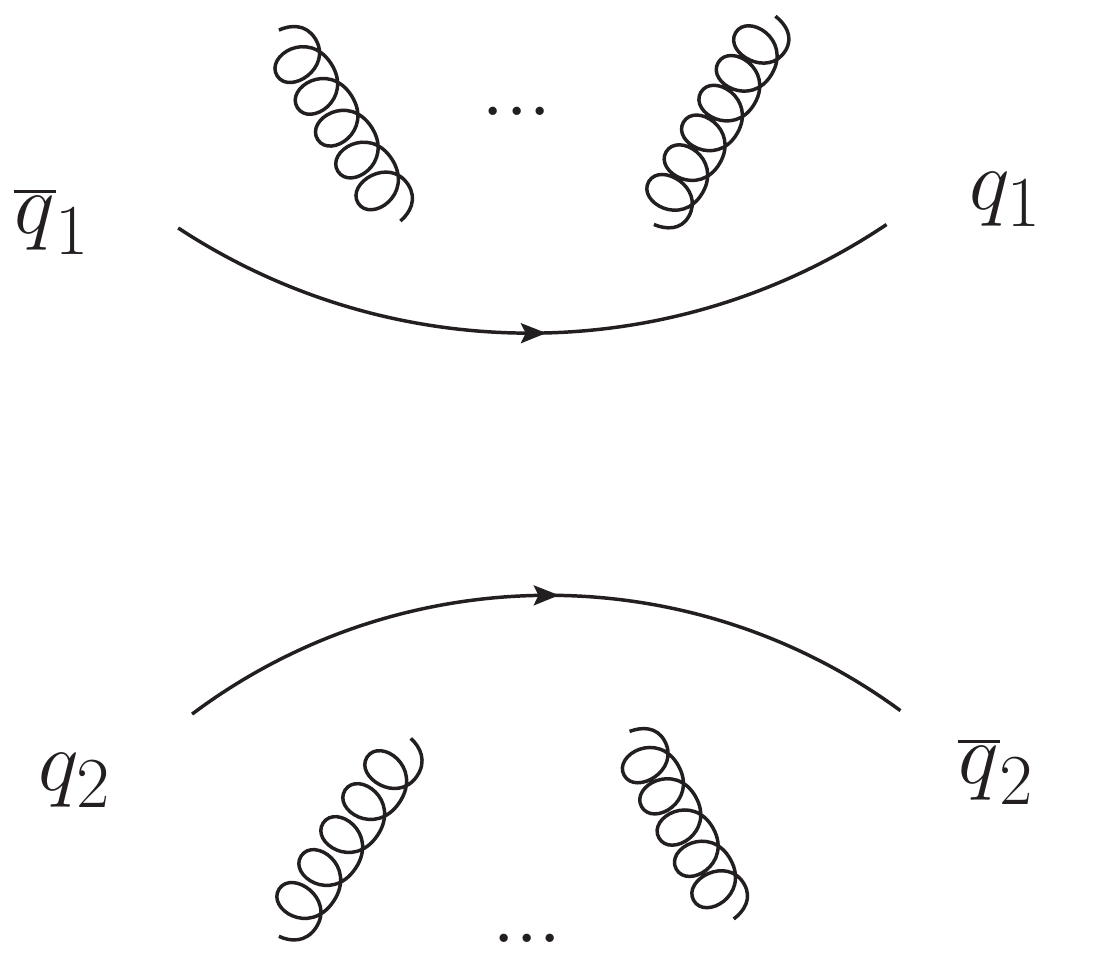}}}
\end{equation}
which is a result of using the  Fierz identity, Eq.~\eqref{eq:Fierz}, for the gluon connecting the two quark lines (note that this is only applicable in this compact way at tree-level). The second term is the contribution in which the ordering is such that
the gluons are all in between a $q\overline{q}$ pair with the same
flavour, resulting in two separate fermion cycles and the $1/\NC$
suppression factor, while for the first term the anti-quarks are
cross-ordered.  We introduce the following notation for these two
contributions
\begin{align}\label{eq:2qq_gen_M}
\M_{2qq}=\M_1-\frac{1}{\NC}\M_2 \qquad \textrm{and}\qquad \M_{2qq}^*=\M_1^*-\frac{1}{\NC}\M_2^*,
\end{align}
for the amplitude and complex conjugate amplitude, respectively, where
\begin{align}\label{eq:2qq_gen_M_2}
  \M_1=g^{n+2}\sum_{\sigma \in S_{n+1}} c_1(\sigma)\A_1(\sigma) \qquad\textrm{and}\qquad
  \M_2=g^{n+2}\sum_{\sigma \in S_{n+1}} c_2(\sigma)\A_2(\sigma)
\end{align}
and similarly for the complex conjugate amplitudes.  In the
expressions for $\M_1$ and $\M_2$ the sum over $\sigma \in S_{n+1}$ is
a sum over all the gluon permutations and the $\overline{q}_2q_2$ (or
$\overline{q}_1q_2$) quark pair, such that the dual amplitudes are
\begin{eqnarray}\label{eq:2qq_amp_types}
\begin{split}
  \A_1(\sigma)&=\A_{2qq}(q_1,\sigma(1),\ldots ,\sigma(n_1+1), \ldots,\sigma(n+1),\overline{q}_1),\\
  \A_2(\sigma)&=\A_{2qq}(q_1,\sigma(1),\ldots , \sigma(n_1+1), \ldots,\sigma(n+1),\overline{q}_2),
  \end{split}
\end{eqnarray}
where the object $\sigma(n_1+1)$ is always the quark-pair index
$\overline{q}_2q_2$ ($\overline{q}_1q_2$) in the case of
$\mathcal{M}_1$ ($\mathcal{M}_2$), and therefore $n_1$ denotes the
number of gluon indices in the permutation $\sigma$ before the
quark-pair index.

The colour coefficients in the fundamental
representation read
\begin{align}\label{eq:colour_two}
\begin{split}
c_1(\sigma)&=(\sA)_{i_1j_2}(\sB)_{i_2j_1}= \left( T^{a_{\sigma(1)}}...T^{a_{\sigma(n_1)}}   \right)_{i_1j_2}\left( T^{a_{\sigma(n_1+1)}}...T^{a_{\sigma(n)}}   \right)_{i_2j_1}, \\
c_2(\sigma)&=(\sC)_{i_1j_1}(\sD)_{i_2j_2} = \left( T^{a_{\sigma(1)}}...T^{a_{\sigma(n_1)}}   \right)_{i_1j_1}\left( T^{a_{\sigma(n_1+1)}}...T^{a_{\sigma(n)}}   \right)_{i_2j_2}
\end{split}
\end{align}
where we introduce the strings of fundamental generators $\sA$ and $\sC$
to correspond to the gluon indices in the permutation $\sigma$
\emph{before} the quark-pair index and $\sB$ and $\sD$ for the ones coming
\emph{after} the quark-pair index, respectively for the $\M_1$ and $\M_2$ amplitudes.

Hence, we obtain the colour-summed squared amplitude, in block matrix
notation,
\begin{multline}
|\M_{2qq}|^2 = (g^2)^{n+2}\sum_{\sigma_k,\sigma_l\in S_{n+1}}\\
\begin{pmatrix}
\A_1(\sigma_k) & \A_2(\sigma_k)
\end{pmatrix}
\begin{pmatrix}
c_1(\sigma_k)c_1(\sigma_l)^*  & -c_1(\sigma_k)c_2(\sigma_l)^*\!\!/\NC \\
-c_2(\sigma_k)c_1(\sigma_l)^*\!\!/\NC  & c_2(\sigma_k)c_2(\sigma_l)^*\!\!/\NC^2 
\end{pmatrix}
\begin{pmatrix}
\A_1(\sigma_l)^* \\
\A_2(\sigma_l)^*
\end{pmatrix}.
\end{multline}
The colour matrix using the notation introduced in Eq.~\eqref{eq:colour_two} can be evaluated to
\begin{equation}\label{eq:2qq_fund}
\renewcommand\arraystretch{1.8}
\mleft(
\begin{array}{c c}
  \Tr[\sA_k\sAt_l]\Tr[\sB_k\sBt_l] & -\Tr[\sA_k\sDt_l\sB_k\sCt_l]/\NC \\
  -\Tr[\sC_k\sBt_l\sD_k\sAt_l]/\NC &  \Tr[\sC_k\sCt_l]\Tr[\sD_k\sDt_l]/\NC^2
\end{array}
\mright)
\end{equation}
where each of the blocks corresponds to a unique relative ordering of
the anti-quarks in the dual amplitudes and conjugate amplitudes, and
within each block, the rows and columns are the permutations of the
gluon and the quark-pair indices.

\subsubsection*{Leading colour}
By inspection, using the usual rules for reduction of the traces of
fundamental matrices (Eqs.~\eqref{eq:rule1}~and~\eqref{eq:rule2}), we
see that the LC contribution comes solely from the upper-left block,
and then only if $\sA_k=\sA_l$ and $\sB_k=\sB_l$, i.e.~the square of
the dual amplitude with cross-ordered quark lines. The
colour factor is equal to
\begin{eqnarray}
  \Tr[\sA_k\sAt_l] \Tr[\sB_k\sBt_l]  =
  \NC^2 \left( \NC - \frac{1}{\NC}\right)^n=\NC^{n+2}+\mathcal{O}(\NC^{n}).
\end{eqnarray}

\subsubsection*{Next-to-leading colour}
For the next-to-leading order in the colour expansion of the colour matrix in Eq.~\eqref{eq:2qq_fund}, we investigate
each of the four blocks separately. We
start with the lower-right block,
\begin{equation}
\Tr[\sC_k\sCt_l]\Tr[\sD_k\sDt_l]/\NC^2,
\end{equation}
since that is the simplest case to consider. Due to the overall factor
$1/\NC^2$, this term gives a NLC contribution equal to
$+\NC^n+\mathcal{O}(\NC^{n-2})$ if and only if $\sC_k=\sC_l$ and
$\sD_k=\sD_l$. Hence, the colour order of the gluon indices, as well as
the quark index need to be identical in the dual amplitude and the
conjugate amplitude for this term to contribute at NLC accuracy.

The upper-right and lower-left blocks are 
\begin{equation}\label{eq:2nd_and_3rd_block}
  -\Tr[\sA_k\sDt_l\sB_k\sCt_l]/\NC \qquad\textrm{and}\qquad
  -\Tr[\sC_k\sBt_l\sD_k\sAt_l]/\NC,
\end{equation}
respectively. These colour factors result in a NLC contribution if and
only if the gluon ordering in the dual conjugate amplitude is such
that
\begin{equation}
    \sC_l=\sA_k^1\sB_k^2,\,\, \sD_l=\sB_k^1\sA_k^2 \qquad\textrm{and}\qquad
    \sA_l=\sC_k^1\sD_k^2,\,\, \sB_l=\sD_k^1\sC_k^2,
\end{equation}
for the first and second terms of Eq.~\eqref{eq:2nd_and_3rd_block},
respectively. Here, the superscripts $1$ ($2$) denote the first
(second) part of a string after splitting it in two in any possible
way, including one of the parts being the empty string, i.e.~equal to
$\mathbb{1}$. The value of the colour factor is equal to
$-\NC^n+\mathcal{O}(\NC^{n-2})$.

Finally, we consider the upper-left block consisting of the colour
factor
\begin{eqnarray}\label{eq:1st_block}
\Tr[\sA_k\sAt_l] \Tr[\sB_k\sBt_l].
\end{eqnarray}
If each of the sets of generators $(\sA_k,\sA_l)$ and $(\sB_k,\sB_l)$
are disjoint, i.e. none of the generators in the first trace have the
same gluon index as any of the generators in the second trace, then
the only contribution is if one of traces satisfy the same condition
as the one-quark-line case in Sec.~\eqref{sec:fund_1qq}, while the pair
of strings in the other trace are identical. Specifically, if
\begin{align}
  &\sA_k = \sR \sQ_1 \sS \sQ_2 \sP,  \qquad \sA_l = \sR \sQ_2 \sS \sQ_1 \sP \qquad\textrm{and}\qquad \sB_k=\sB_l \qquad\textrm{\emph{or}}\\
  &\sB_k = \sR \sQ_1 \sS \sQ_2 \sP,  \qquad \sB_l = \sR \sQ_2 \sS \sQ_1 \sP \qquad\textrm{and}\qquad \sA_k=\sA_l
\end{align}
this yields an NLC factor, $+\NC^{n}+\mathcal{O}(\NC^{n-2})$, with the
following exceptions.
\begin{itemize}
\item If $\sS=\mathbb{1}$ and $\len(\sQ_1)=\len(\sQ_2)=1$ the colour factor is a
  NLC term equal to $-\NC^{n}+\mathcal{O}(\NC^{n-2})$;
\item If $\sS=\mathbb{1}$ and $\len(\sQ_1)=1$ \emph{or} $\len(\sQ_2)=1$ the colour factor is not a NLC term, $\mathcal{O}(\NC^{n-2})$.
\end{itemize}
In the case that both $\sA_k=\sA_l$ and $\sB_k=\sB_l$, this term contributes
at LC, but also at NLC; to be specific, $\Tr[\sA_k\sAt_l]
\Tr[\sB_k\sBt_l]=\NC^{n+2}-n\NC^{n}+\mathcal{O}(\NC^{n-2})$.

If the sets of generators are non-disjoint,
i.e.~$(\sA_k,\sAt_l)\cap(\sB_k,\sBt_l) \neq \emptyset$, the situation
is more involved. In this case, we then know that at least one
generator in $\sA_k$ has the same gluon index as a generator in
$\sB_l$ (or, equivalently, there is a common generator between $\sA_l$
and $\sB_k$), thus we can write, without loss of generality,
\begin{align}\label{eq:non-disjoint}
\sA_k=\sV_kT^a \sW_k \qquad\textrm{and}\qquad \sB_l=\sI_l T^a \sJ_l,
\end{align}
where, if there is more than one common generator, the $T^a$ picked
here is the first one appearing in the $\sA_k$ string, i.e.~all
generators in $\sV_k$ are contracted with generators in $\sAt_l$,
while this is not necessarily the case for the $\sW_k$.  It is
possible that the $\sV_k$, $\sW_k$, $\sI_l$, and/or $\sJ_l$ strings
are the identity. The colour factor in Eq.~\eqref{eq:1st_block}
contributes to NLC as $+\NC^n+\mathcal{O}(\NC^{n-2})$ if
\begin{align}\label{eq:2qq_disjoint_res}
  \sA_l^1=\sV_k\qquad\textrm{and}\qquad
  \sJ_l=\sW_k^1\sB_k^2\qquad\textrm{and}\qquad
  \sI_l\sA_l^2=\sB_k^1\sW_k^2,
\end{align}
where, as before, the superscripts $1$ ($2$) denote the first (second)
part of a string after splitting it in two in any possible way,
including one of the parts being an empty string. There is one
exception to this rule, and that is that this does not result in a NLC
term if $\sA_l = \sV_k \sW_k$ or $\sB_k=\sI_l \sJ_l$.

\subsubsection*{Proof of the NLC terms}
The lower-right block in Eq.~\eqref{eq:2qq_fund} is straight-forward
and does not require any proof. For the upper-right and lower-left
blocks we first realise that these colour factors can yield a maximum
colour factor of $\mathcal{O}(\NC^{n})$ and that is obtained only if
all indices are contracted at once without having to rewrite into a
product of traces. In other words, it must be possible to simplify the
traces using only Rule~IIb, Eq.~\eqref{eq:rule2b}, which
means that at any point two neighbouring generators can be contracted.
Let us consider the first term in Eq.~\eqref{eq:2nd_and_3rd_block};
the second term can be dealt with in an equivalent manner. Let the
length of the $\sA_k$ and $\sB_k$ be $a$ and $b$, respectively. We can
split these strings of fundamental generators into two strings,
$\sA_k=\sA_k^{1}\sA_k^{2}$ and $\sB_k=\sB_k^{1}\sB_k^{2}$, such that
$\len(\sA_k^{i})=a_i$ and $\len(\sB_k^{i})=b_i$, $i=1,2$. We have $a_2=a-a_1$ and
$b_2=b-b_1$. The colour factor given by
$-\Tr[\sA_k\sDt_l\sB_k\sCt_l]/\NC$ results in a NLC factor if
and only if
\begin{equation}
  \sCt_l=\sBt_k^{2}\sAt_k^{1} \quad\textrm{and}\quad
  \sDt_l=\sAt_k^{2}\sBt_k^{1}
\end{equation}
for any $a_1\in\{0,\ldots,a\}$ and $b_1\in\{0,\dots,b\}$.  With the
gluons in the dual conjugate amplitude ordered in this way, there are
always neighbouring generators that are identical in the first term of
Eq.~\eqref{eq:2nd_and_3rd_block} and can be contracted using Rule~IIb.
The same argument applies to the $-\Tr[\sC_k\sBt_l\sD_k\sAt_l]/\NC$
colour factor, i.e.~the second term of
Eq.~\eqref{eq:2nd_and_3rd_block}.

The proof for the disjoint case for the upper-left block in
Eq.~\eqref{eq:2qq_fund} follows directly from the proof for the
one-quark-line case, Sec.~\eqref{sec:fund_1qq} and we will not repeat
it here. For the non-disjoint situation, we have the configuration as
stated in Eq.~\eqref{eq:non-disjoint}. Therefore, we can use Rule~I,
Eq.~\eqref{eq:rule1}, to simplify the colour factor, which results in
\begin{eqnarray}\label{eq:trace_1}
  \Tr[\sW_k\sAt_l\sV_k \sIt_l\sB_k\sJt_l]-
  \frac{1}{\NC}\Tr[\sW_k\sAt_l\sV_k]\Tr[\sIt_l\sB_k\sJt_l].
\end{eqnarray} 
Since, we have now applied the Rule~I once, for these colour factors
to contribute at NLC, it must be possible to simplify them completely
using only Rule~IIb, Eq.~\eqref{eq:rule2b}. Since all the
generators in $\sV_k$ are contracted with generators in
$\sAt_l$ by construction, it must be that the $\sA_l$ string starts
with the same generators (and they must be in the same order) as
$\sV_k$ for the first term in Eq.~\eqref{eq:trace_1} to
potentially contribute at NLC. What is left is a string of the same
structure as dealt with in Eq.~\eqref{eq:2nd_and_3rd_block}, and a
similar derivation follows, which we will not repeat here, and results
in Eq.~\eqref{eq:2qq_disjoint_res}. The second term in
Eq.~\eqref{eq:trace_1} can also contribute at NLC, and with opposite
sign, canceling the first term. For this to
be the case, the two separate traces should simplify using only
Rule~IIb. Hence, if $\sA_l = \sV_k \sW_k$ or
$\sB_k=\sI_l \sJ_l$ the $\Tr[\sA_k\sAt_l]
\Tr[\sB_k\sBt_l]$ term in the colour matrix vanishes.

\subsection{Two same-flavour $q\overline{q}$ pairs and $n$ gluons}\label{sec:fund_2qq_SF}
The colour-summed squared amplitudes for the two same-flavour quark lines can be
obtained from the distinct-flavour case by symmetrising the result. In
particular, we can expand the amplitude into
\begin{eqnarray}\label{eq:2qq_sf}
  \M_{2qq}(\overline{q}q\overline{q}q+n g) =
  \hat{\M}(\overline{q}_1 q_1 \overline{q}_2 q_2+ng) -
  \hat{\M}(\overline{q}_1 q_2 \overline{q}_2 q_1+ng) 
\end{eqnarray}
where the amplitudes with hats denote amplitudes in which the first
two and last two quark indices are two distinct-flavoured pairs of
quarks and with the relative minus sign between these two amplitudes arising from Fermi statistics. As in Eq.~\eqref{eq:2qq_gen_M}, the amplitudes can be
decomposed in two orderings for the quark indices
\begin{align}
  \hat{\M}(\overline{q}_1 q_1 \overline{q}_2 q_2+ng) &= \mathcal{M}_1 - \frac{1}{\NC} \mathcal{M}_2, \\
  \hat{\M}(\overline{q}_1 q_2 \overline{q}_2 q_1+ng) &= \mathcal{M}_2 - \frac{1}{\NC} \mathcal{M}_1,
\end{align}
where $\M_1$ and $\M_2$ are defined in Eq.~\eqref{eq:2qq_gen_M_2}. This
yields,
\begin{eqnarray}
\M_{2qq} = \left( 1+ \frac{1}{\NC} \right) ( \M_1 - \M_2 ).
\end{eqnarray}
Note that now, as opposed to the distinct-flavour case, the dual amplitudes appearing in $\M_1$ and $M_2$ are put on equal footing.
Using the expressions for $\M_{1,2}$ from Eq.~\eqref{eq:2qq_gen_M_2}, we obtain the colour-summed squared amplitude in block matrix form
\begin{multline}\label{eq:2qq_fund_distinct}
|\M_{2qq}|^2 = (g^2)^{n+2}\left(1+\frac{1}{\NC}\right) ^2\sum_{\sigma_k,\sigma_l\in S_{n+1}} \\
\begin{pmatrix}
  \A_1(\sigma_k) & \A_2(\sigma_k)
\end{pmatrix}
\begin{pmatrix}
  c_1(\sigma_k) c_1(\sigma_l)^*  & -c_1(\sigma_k)c_2(\sigma_l)^* \\
  -c_2(\sigma_k) c_1(\sigma_l)^*  & c_2(\sigma_k)c_2(\sigma_l)^* 
\end{pmatrix}
\begin{pmatrix}
  \A_1(\sigma_l)^* \\
  \A_2(\sigma_l)^*
\end{pmatrix}.
\end{multline}
The colour matrix has the block structure in terms of the amplitudes
and corresponding colour factor structures, using the notation introduced in Eq.~\eqref{eq:colour_two},
\begin{equation}\label{eq:2qq_SF_blocks}
\renewcommand\arraystretch{1.8}
\mleft(
\begin{array}{cc}
  \Tr[\sA_k\sAt_l] \Tr[\sB_k\sBt_l] & -\Tr[\sA_k\sDt_l\sB_k\sCt_l] \\
  -\Tr[\sC_k\sBt_l\sD_k\sAt_l]      & \Tr[\sC_k\sCt_l] \Tr[\sD_k\sDt_l]
\end{array}
\mright)
\end{equation}
where within each block, the rows and columns are the permutations of
the gluon indices and the quark-pair index. 

Contrary to all the other cases considered so far, for the two
same-flavour $q\overline{q}$ pairs and $n$ gluons, the colour
expansion is not an expansion in $1/\NC^2$, but rather an expansion in
$1/\NC$. The LC contribution is of order
$\mathcal{O}(\NC^{n+2})$, while we define the NLC to include both
$\mathcal{O}(\NC^{n+1})$ and $\mathcal{O}(\NC^{n})$. Terms of order
$\mathcal{O}(\NC^{n-1})$ and beyond will be neglected in our expansion
up to NLC accuracy, similarly to the distinct-flavour case.

\subsubsection*{Leading colour}
Two blocks in the colour matrix can give LC contributions: if
$\sA_k=\sA_l$ and $\sB_k=\sB_l$ (for the upper-left block) or $\sC_k=\sC_l$ and
$\sD_k=\sD_l$ (for the lower-right block). The colour factors in these cases are
\begin{eqnarray}\label{eq:2qq_sf_lc}
\begin{split}
  \left(1+\frac{1}{\NC}\right)^2\Tr[\sA_k\sAt_l] \Tr[\sB_k\sBt_l] \quad  \text{or} \quad   \left(1+\frac{1}{\NC}\right)^2\Tr[\sC_k\sCt_l] \Tr[\sD_k\sDt_l]  \\
  =
  \left(1+\frac{1}{\NC}\right)^2 \NC^2 \left( \NC - \frac{1}{\NC}\right)^n=\NC^{n+2}+\mathcal{O}(\NC^{n+1}).
  \end{split}
\end{eqnarray}
The difference from the distinct-flavour case arises because the lower-right block now also contributes to LC, as opposed to the distinct-flavour case, where this block only contributes at NLC.

\subsubsection*{Next-to-leading colour}
To determine the NLC contributions in the upper-left and lower-right
blocks of the colour matrix, the arguments developed for the
upper-left block in the distinct-flavour case apply here to both
these blocks, with the one subtlety that the LC contribution
contributes differently to NLC due to the $(1+1/\NC)^2$ prefactor in
Eq.~\eqref{eq:2qq_sf_lc}; to be specific
$(1+1/\NC)^2\Tr[\sA_k\sAt_l]\Tr[\sB_k\sBt_l]=\NC^{n+2}+2
\NC^{n+1}-(n-1)\NC^{n}+\mathcal{O}(\NC^{n-2})$. Also the off-diagonal
blocks of the colour matrix have the same form as in the
distinct-flavour case. The difference is that in the distinct-flavour
case, they come with a $-1/\NC$ prefactor, while here the prefactor is
less-suppressed, $-(1+1/\NC)^2$. Since the difference is only one
power of $\NC$, this does not mandate the need for keeping more terms
in the expansion in the computation of the trace, and all arguments
developed for the distinct-flavour quark lines also apply here, with
the only difference that the value of the colour factor is now equal
to $-\NC^{n+1}+2\NC^{n}+\mathcal{O}(\NC^{n-1})$.

\section{The colour matrix in the colour-flow decomposition}\label{sec:CF}
In the colour-flow decomposition, the colour factors are strings of
Kronecker deltas, with one index in the fundamental representation and
one in the anti-representation. As such, this basis is a physically
intuitive basis, in which the colour factor is directly given by
connecting the colour-flow lines. In this decomposition, external $\SU(\NC)$ gluons
are projected onto a $\U(\NC)$ part and $\U(1)$ part and are treated
as different particles when constructing the dual amplitudes.

The colour factors are obtained by contracting the strings of
Kronecker deltas. A closure of the indices is denoted as a colour loop
(or fermion loop) and yields a power of $\NC$. Thus, a pure
contraction of the strings yields monomials in $\NC$, with the power
equal to the number of closed loops in the string. Below we present
the (N)LC contributions for the all-gluon, one-quark-line plus gluons
and two-quark-line plus gluons (both distinct-flavour and
same-flavour cases) amplitudes.

\subsection{$n$-gluon amplitude}\label{sec:gluon_CF}
In the colour-flow decomposition for the all-gluon amplitude, the
squared amplitude, summed over all colours, matches that of
Eq.~\eqref{eq:squared_gluons} with the colour
matrix~\cite{Maltoni:2002mq} taking the form
\begin{align}
  C(\sigma_k,\sigma_l) =& \delta^{i_n}_{j_{\sigma_k(1)}}\delta^{j_{\sigma_k(1)}}_{i_{\sigma_k(2)}}\ldots \delta^{i_{\sigma_k(n-1)}}_{j_n}
            \left( \delta^{i_n}_{j_{\sigma_l(1)}}\delta^{j_{\sigma_l(1)}}_{i_{\sigma_l(2)}}\ldots \delta^{i_{\sigma_l(n-1)}}_{j_n}  \right)^{\dagger} \nonumber\\
            =& \delta^{i_n}_{j_{\sigma_k(1)}}\delta^{j_{\sigma_k(1)}}_{i_{\sigma_k(2)}}\ldots \delta^{i_{\sigma_k(n-1)}}_{j_n}
            \left( \delta_{i_n}^{j_{\sigma_l(1)}}\delta_{j_{\sigma_l(1)}}^{i_{\sigma_l(2)}}\ldots \delta_{i_{\sigma_l(n-1)}}^{j_n}  \right),\label{eq:kronecker}
\end{align}
with a sum over all gluon pair of indices $(i,j)$. 

\subsubsection*{Leading colour}
The leading term in the large-$\NC$ expansion is obtained for
$\sigma_k=\sigma_l$, for which the colour factor is equal to $\NC^n$.

\subsubsection*{Next-to-leading colour}
The NLC interference contributions, for which the colour factor is
$\NC^{n-2}$, in the colour matrix are very similar to that for the
fundamental decomposition, but now without the list of
exceptions. Hence, the NLC come from, and only from, a difference of a
single permutation of a string of Kronecker deltas between the colour
ordering in the dual amplitude and conjugate amplitudes. That is, schematically,
\begin{eqnarray}\label{eq:color-flow-gluon}
\sigma_R \sigma_{Q_1} \sigma_S \sigma_{Q_2} \sigma_P  \times \left(\sigma_R \sigma_{Q_2} \sigma_S \sigma_{Q_1} \sigma_P\right)^{\dagger}
\end{eqnarray}
where all $\sigma_{R,S,P,Q_{1,2}}$ are subpermutations of $\sigma_{k,l}$
with specific ordering of gluon indices. Note that all the
subpermutations $\sigma$ in the above may be empty, apart from
$\sigma_{Q_1,Q_2}$, where each need to contain at least one element.

\subsubsection*{Proof of the NLC terms}
We consider the interference between two amplitudes, with the colour
factor expressed in the colour-flow decomposition. In this basis, the
factor is a set of $2n$ Kronecker deltas, as given in
Eq.~\eqref{eq:kronecker} with all gluon indices where each of the two
indices of the external gluons appear twice in the string of
deltas. Contracting any two Kronecker deltas which contain a certain
index in its representation and anti-representation, two outcomes are
possible: either it yields a factor of $\NC$ or a new Kronecker delta
(depending on what the other index in the Kronecker deltas is). The
resulting string is thus of length $2n-2$ or $2n-1$, and with a power
of $N_C^1$ or $N_C^0$. This series of contractions is repeated until
all gluon indices are contracted in a first round and one is left with
a new set of Kronecker deltas (or none if all indices are contracted
in the first round). The indices left are then contracted in a similar
fashion in a second round. This is repeated until no Kronecker deltas
are left and are replaced by monomials of $\NC$. Following through all
possible contractions, one finds that the leading-colour is obtained
when each Kronecker delta contraction yields only $N_C$, resulting in
a colour factor of order $\NC^n$. The next order in the colour
expansion (NLC) is found to be $\NC^{n-2}$, which is obtained if only
two pairs of gluon indices yield a new Kronecker delta at contraction
in the first round, while all other contractions yield a factor
$N_C$. This is the condition which is stated in
Eq.~\eqref{eq:color-flow-gluon} and thus completes the proof.

\subsection{One $q\overline{q}$ pair and $n$ gluons}\label{sec:color-flow-1quark}
For a single quark pair in the colour-flow decomposition, the colour
decomposition becomes somewhat more extensive due to the possibility
of external $\U(1)$ gluons coupling to the quark line. The colour
factors for these contributions are however directly obtained from the
case of $\U(\NC)$ gluon emissions, by the replacement
\begin{eqnarray}\label{eq:replace}
  \delta^{i}_j\delta^{k}_{l} \rightarrow \delta^{i}_{j}\delta^{k}_{l} -
  \frac{1}{\NC}  \delta^{i}_{l}\delta^{k}_{j} 
\end{eqnarray}
as explained in more detail in Ref.~\cite{Maltoni:2002mq}. Applying
this to all the external $n$ gluons finally yields the following
expression for the amplitude~\cite{Maltoni:2002mq,Hagiwara:2010vk}
\begin{align}\label{eq:UN_and_U1-gluons}
\M_{1qq} = & g^n\sum_{\sigma\in S_n} \delta^{i_q}_{j_{\sigma(1)}}\delta^{i_{\sigma(1)}}_{j_{\sigma(2)}}\ldots  \delta^{i_{\sigma(n-1)}}_{j_{\sigma(n)}}\delta^{i_{\sigma(n)}}_{j_q}\A_{1qq}(q,\sigma(1),\ldots ,\sigma(n),\overline{q} ) \nonumber\\
&+\left(\frac{-1}{\NC}\right)g^n\sum_{\sigma\in S_n} \delta^{i_q}_{j_{\sigma(1)}}\delta^{i_{\sigma(1)}}_{j_{\sigma(2)}}\ldots  \delta^{i_{\sigma(n-1)}}_{j_q}\delta^{i_{\sigma(n)}}_{j_{\sigma(n)}}\A_{1qq}(q,\sigma(1),\ldots ,\sigma(n-1),\overline{q},\sigma(n) ) \nonumber\\
&+\left(\frac{-1}{\NC}\right)^2 \frac{1}{2!}g^n\sum_{\sigma\in S_n} \delta^{i_q}_{j_{\sigma(1)}}\delta^{i_{\sigma(1)}}_{j_{\sigma(2)}}\ldots  \delta^{i_{\sigma(n-2)}}_{j_{q}}\delta^{i_{\sigma(n-1)}}_{j_{\sigma(n-1)}}\delta^{i_{\sigma(n)}}_{j_{\sigma(n)}}\A_{1qq}(q,\sigma(1),\ldots ,\sigma(n-2),\overline{q},\sigma(n-1),\sigma(n) ) \nonumber\\
&+\,\ldots  \nonumber\\
&+ \left(\frac{-1}{\NC}\right)^n \frac{1}{n!}g^n\sum_{\sigma\in S_n} \delta^{i_q}_{j_{q}}\delta^{i_{\sigma(1)}}_{j_{\sigma(1)}}\ldots  \delta^{i_{\sigma(n-1)}}_{j_{\sigma(n-1)}}\delta^{i_{\sigma(n)}}_{j_{\sigma(n)}}\A_{1qq}(q,\overline{q},\sigma(1),\ldots ,\sigma(n) ).
\end{align}
The notation for the dual amplitudes is that the gluon indices after
the anti-quark label denote the $\U(1)$ external gluon indices, and
the value of these amplitudes is independent from the order of these
gluons, which is the reason for the $1/r!$ factorial factors, with $r$
the number of $\U(1)$ gluons in the dual amplitude. The dual
amplitudes with external $\U(1)$ gluons are however not independent,
but can be expressed as linear combinations of the ones with external
$\U(\NC)$ gluons \cite{Maltoni:2002mq,Frixione:2021yim}. This property
will be utilized to make the colour matrix in the colour-flow
decomposition more sparse.

As highlighted also in Ref.~\cite{Frixione:2021yim}, the expansion in
Eq.~\eqref{eq:UN_and_U1-gluons} is in terms of dual amplitudes with
unphysical gluons, as $\U(\NC)$ and $\U(1)$ counterparts of the
physical $\SU(\NC)$ gluons. However, each term is significant in order
to render the sum physical. The space of $\SU(\NC)$ (physical) gluons
and that of the $\U(1)$ counterparts are however orthogonal, and
hence, the interference terms between a $\U(\NC)$ and a $\U(1)$ gluon
projects onto an interference between the $\U(1)$ part of the
$\U(\NC)$ and the $\U(1)$ gluon. Nevertheless, one needs to consider
each contribution in the colour matrix in order to perform the colour
sum.

\subsubsection*{Leading colour}
As the number of Kronecker deltas remains the same in each of the terms
in Eq.~\eqref{eq:UN_and_U1-gluons} for different number of external
$\U(1)$ gluons, it is clear that only the first term squared can give
LC contribution, for which all Kronecker deltas in the squared
amplitude contract and give a factor of $\NC$. That means that the
permutations $\sigma_k$ in the amplitude and $\sigma_l$ in the conjugate
amplitude need to be the same. The LC colour factor is then
given by
\begin{equation}
  \delta^{i_q}_{j_{\sigma_k(1)}}\delta^{i_{\sigma_k(1)}}_{j_{\sigma_k(2)}}\ldots  \delta^{i_{\sigma_k(n-1)}}_{j_{\sigma_k(n)}}\delta^{i_{\sigma_k(n)}}_{j_q}  \left(  \delta^{i_q}_{j_{\sigma_l(1)}}\delta^{i_{\sigma_l(1)}}_{j_{\sigma_l(2)}}\ldots  \delta^{i_{\sigma_l(n-1)}}_{j_{\sigma_l(n)}}\delta^{i_{\sigma_l(n)}}_{j_q} \right)^{\dagger} = \NC^{n+1},
\end{equation}
with $\sigma_k=\sigma_l$.

\subsubsection*{Next-to-leading colour}
The interference terms between the dual amplitudes appearing in
Eq.~\eqref{eq:UN_and_U1-gluons} can be further simplified by, as
already noted previously, using instead of the dual amplitude with one
$\U(1)$ gluon, the linear combination of dual amplitudes with only
$\U(\NC)$ gluons. More precisely,
\begin{eqnarray}\label{eq:sum_U1}
\begin{split}
\A(q,\sigma(1),\ldots,\sigma(n-1),\overline{q},\sigma(n)) = &\A(q,\sigma(n),\sigma(1),\ldots,\sigma(n-1),\overline{q}) + \\
&\A(q,\sigma(1),\sigma(n),\ldots,\sigma(n-1),\overline{q}) + \\ & ... \\
&\A(q,\sigma(1),\ldots,\sigma(n-1),\sigma(n),\overline{q}).
\end{split}
\end{eqnarray}
It is however evident that upon replacing each of the amplitudes with
$\U(1)$ gluons of any number, the final result of the set of dual
amplitudes and their corresponding colour factors is precisely that of
the fundamental decomposition. We will therefore utilize this
simplification only to some extent, which will be made clear under the
proof section. In particular, our colour factors will remain monomials
in $\NC$.
 
NLC colour factors arise in the interferences of two types of dual
amplitudes. Firstly, the NLC elements in the colour matrix arise in
the rows with a dual amplitude with only external $\U(\NC)$ gluons,
with conjugate amplitudes with also purely $\U(\NC)$ gluons,
effectively taking the first line of Eq.~\eqref{eq:UN_and_U1-gluons}
for both the amplitude and the conjugate amplitude. This situation is
the same as the all-gluon case, with NLC elements if and only if the
permutations of the dual amplitude and conjugate amplitude satisfy the
block-interchange-relation
\begin{align}\label{eq:1qq_pos1_CF}
\sigma_R \sigma_{Q_1} \sigma_S \sigma_{Q_2} \sigma_P \times \left(\sigma_R \sigma_{Q_2} \sigma_S \sigma_{Q_1} \sigma_P\right)^{\dagger}.
\end{align}

Secondly, NLC accurate colour factors are also in the rows for dual
amplitudes with one external $\U(1)$, interfering with conjugate
amplitudes with the same permutation $\sigma_k=\sigma_l$, with a
modified entry in the colour matrix of $-n\NC^{n-1}$.

\subsubsection*{Proof of the NLC terms}
The proof for the first possibility, i.e.~no $\U(1)$ gluons, follows
the proof for the all-gluon case with the $q\overline{q}$ pair playing
the role of first gluon, i.e.~the gluon that is not permuted, and we
will therefore not repeat it here.

\begin{figure}[b!]
\center
\includegraphics[scale=0.3]{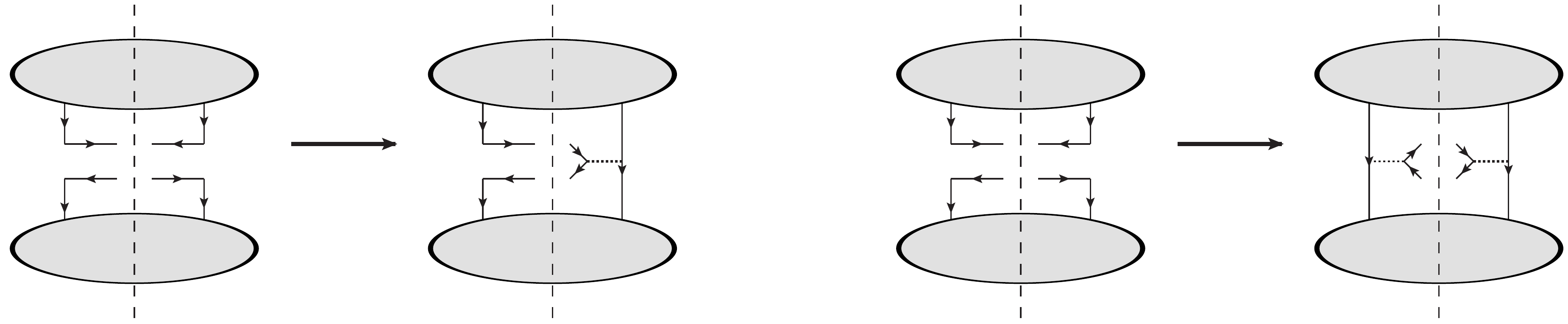}
\caption{Contributions to the matrix element squared, in which in the
  left (right) two diagrams one (both) $\U(\NC)$ gluon(s) is (are)
  replaced by $\U(1)$ gluon(s) on the left or (and) right of the
  unitarity cut.}
\label{fig:gluons}
\end{figure}

For the cases where we include external $\U(1)$ gluons, we use a
diagrammatic approach. For each closed colour loop, we have a power of
$\NC$ and each $\U(1)$ gluon comes with a factor of $-1/\NC$.  Hence,
we only need to consider the cases with up to two external $\U(1)$
gluons in total in the amplitude and conjugate amplitude.

When introducing external $\U(1)$ gluons, there can be two types of
connections between the amplitude and conjugate amplitude: one in
which the $\U(1)$ gluon in the dual amplitude connects to a $\U(\NC)$
gluon in the conjugate amplitude (or vice versa), or a $\U(1)$ gluon
connects with a $\U(1)$ gluon, see the left and right sets of diagrams
in Fig.~\ref{fig:gluons}, respectively. The possibility depicted on
the left of this figure reduces the number of total closed colour
loops by one and introduces one factor of $-1/\NC$, in total reducing
the power of $\NC$ by 2. For the right part of Fig.~\ref{fig:gluons},
the total number of loops is conserved, while two factors of $-1/\NC$
are introduced. Hence, also in this case there is a total reduction of
powers of $\NC$ by 2. Hence, these two possibilities give NLC
contributions if all the other Kronecker deltas give the maximum
allowed number of colour loops. That means that the colour ordering
must be the same in the amplitude and the conjugate amplitude. This
results precisely in the two cases presented. In the following we
describe how to use Eq.~\eqref{eq:sum_U1} to further reduce the number
of conjugate amplitude per dual amplitude.

We consider the possibility of an interference with one external
$\U(1)$ gluon in \emph{both} the amplitude and the conjugate
amplitude, an interference with both amplitude and conjugate amplitude
of the type in the second line in Eq.~\eqref{eq:UN_and_U1-gluons}. For
this to contribute at NLC, an NLC factor of $\NC^{n-1}$, the same
gluon in the amplitude and the conjugate amplitude must be the
$\U(1)$~gluon, and the order of the other gluons must be identical, so
$\sigma_k=\sigma_l$ for the permutation of the amplitude and conjugate
amplitude respectively, where it is understood that the last element
of the permutation denotes the $\U(1)$ gluon. This is a single
conjugate amplitude for each dual amplitude of this type yielding a
NLC entry in the colour matrix.

Next we consider the interferences where the dual amplitude is with
one external $\U(1)$ gluon and the conjugate amplitude with only
$\U(\NC)$ gluons. For this to be a NLC contribution of colour factor
$-\NC^{n-1}$, the colour order of the $n-1$ $\U(\NC)$~gluons in the
amplitude must be identical to the order of these gluons in the
conjugate amplitude, while the $\U(1)$ indices might be any of the
$\U(\NC)$~gluon indices in the conjugate amplitude. More precisely, if
$\sigma_k$ denotes the permutation of the amplitude with one external
$\U(1)$ gluon where the final element is always understood as the
external $\U(1)$ gluon, then its interference with the permutation
$\sigma_l$ which preserves the relative ordering of the $n-1$ first
elements of $\sigma_k$, but the $\sigma_k(n)$ index is located in any
position in $\sigma_l$, yields a NLC factor. This is a total of $n$
conjugate amplitudes to consider for each dual amplitude with one
external $\U(1)$ gluon. Hence, the sum of these interferences can be
written as
\begin{eqnarray}
\begin{split}
-\NC^{n-1} \A_{1qq}(q,\sigma_k(1),\ldots,\sigma_k(n-1),\overline{q},\sigma_k(n)) \times & \left [ (\A_{1qq}(q,\sigma_k(n),\sigma_k(1),\ldots,\sigma_k(n-1),\overline{q}))^*+ \right. \\
& \left. (\A_{1qq}(q,\sigma_k(1),\sigma_k(n),\ldots,\sigma_k(n-1),\overline{q}))^*+ \right. \\
& \left. \ldots \right. \\
& \left. (\A_{1qq}(q,\sigma_k(1),\ldots,\sigma_k(n-1),\sigma_k(n),\overline{q}))^* \right]
\end{split}
\end{eqnarray}
where we note that the sum in the square brackets is precisely the linear combination of the one $\U(1)$-type amplitude $\A_{1qq}(q,\sigma_k(1)...\sigma_k(n-1),q,\sigma_k(n))^*$ as given also in Eq.~\eqref{eq:sum_U1}. Therefore, this contribution cancels precisely the only interference which appears in the colour matrix between a $\U(1)$ and $\U(1)$ type amplitude with colour factor $+\NC^{n-1}$ which was discussed in the previous paragraph. As such, these contributions may freely be omitted completely from the colour matrix without changing the colour sum.

Finally, there is the interference between a dual amplitude with all $\U(\NC)$ gluons and a conjugate amplitude with one $\U(1)$ gluon. The possible conjugate amplitudes yielding a NLC are those for which any of the gluons  in the amplitude are a $\U(1)$ in the conjugate amplitude, but the relative ordering of the remaining $\U(\NC)$ gluons are unchanged.  This results in $n$ possible conjugate amplitude to each dual amplitude of this type. These interferences appear with a factor of $-\NC^{n-1}$, of the form
\begin{eqnarray}\label{eq:U1_proj}
\begin{split}
-\NC^{n-1} \A_{1qq}(q,\sigma_k(1),\ldots,\sigma_k(n),\overline{q}) \times & \left [ (\A_{1qq}(q,\sigma_k(2),\sigma_k(3),\ldots,\sigma_k(n),\overline{q},\sigma_k(1)))^*+ \right. \\
& \left. (\A_{1qq}(q,\sigma_k(1),\sigma_k(3),\ldots,\sigma_k(n),\overline{q},\sigma_k(2)))^*+ \right. \\
& \left. \ldots \right. \\
& \left. (\A_{1qq}(q,\sigma_k(1),\sigma_k(2),\ldots,\sigma_k(n-1),\overline{q},\sigma_k(n)))^* \right]
\end{split}
\end{eqnarray}
where in each interference, the $\U(1)$ part of the $\U(\NC)$ gluon in the dual amplitude with the index which the $\U(1)$ gluon carries in the conjugate amplitude, is projected out, reducing the interference of the type with one external $\U(1)$ gluon in both the dual amplitude and the conjugate amplitude. Thus, these interference terms may be moved to the diagonal part of the colour matrix with one external $\U(1)$ gluon amplitude types. The only subtlety is that the colour factor appearing in these diagonal entries are now $-n\NC^{n-1}$, which accounts for the $n$ rows in the colour matrix with amplitudes with only $\U(\NC)$ gluons which interfere with each of the columns of the colour matrix with a specific gluon ordering and single $\U(1)$ gluon index.

Using the same reasoning we see that an interference type with an
amplitude with only $\U(\NC)$ gluons and a conjugate amplitude with
two external $\U(1)$ gluons, the number of colour loops will be
reduced by two while simultaneously introducing two factors of
$-1/\NC$, hence, this type of interference will not yield a NLC colour
factor. This completes the proof.

\subsection{Two distinct-flavour $q\overline{q}$ pairs and $n$ gluons}
The new feature for amplitudes with two quark lines, is that we must
allow for the possibility that the intermediate gluon between the two
quark lines is a $\U(1)$ gluon. This is similar to the decomposition
of the internal gluon in the fundamental basis in
Eq~\eqref{eq:2qq_decomp} in terms of the Fierz identity. This implies
already an explicit $-1/\NC$ contribution from the internal $\U(1)$
gluon propagator. In particular, the amplitude can be written in the
same manner as in the fundamental decomposition~\cite{Maltoni:2002mq}
\begin{equation}\label{eq:2qq_color}
  \M_{2qq} = \M_1-\frac{1}{\NC}\M_2,
\end{equation}
with the amplitudes
\begin{equation}\label{eq:2qq_color_simple}
  \M_1=g^{n+2}\sum_{\sigma\in S_{n+1}} c_1(\sigma)\A_1(\sigma) \quad\textrm{and}\quad
  \M_2=g^{n+2}\sum_{\sigma\in S_{n+1}} c_2(\sigma)\A_2(\sigma),
\end{equation}
and with the colour factors $c_{1,2}$ and dual amplitudes $\A_{1,2}$
defined as
\begin{eqnarray}\label{eq:2qq_c_and_A}
\begin{split}
  c_1(\sigma)&=\delta^{i_{q_1}}_{j_{\sigma(1)}}\ldots \delta^{i_{\sigma(n_1)}}_{j_{q_2}}\delta^{i_{q_2}}_{j_{\sigma(n_1+1)}}\ldots \delta^{i_{\sigma(n)}}_{j_{q_1}}, \\
  c_2(\sigma)&=\delta^{i_{q_1}}_{j_{\sigma(1)}}\ldots \delta^{i_{\sigma(n_1)}}_{j_{q_1}}\delta^{i_{q_2}}_{j_{\sigma(n_1+1)}}\ldots \delta^{i_{\sigma(n)}}_{j_{q_2}},\\
    \A_1(\sigma)&=\A_{2qq}(q_1,\sigma(1),\ldots ,\sigma(n_1+1), \ldots,\sigma(n+1),\overline{q}_1),\\
  \A_2(\sigma)&=\A_{2qq}(q_1,\sigma(1),\ldots , \sigma(n_1+1), \ldots,\sigma(n+1),\overline{q}_2).
  \end{split}
\end{eqnarray}
Note, in particular, the difference in the order of the anti-quarks in
the colour factors. The dual amplitudes are the same ones appearing in
the fundamental basis in Eq~\eqref{eq:2qq_amp_types}. Similarly to the
case of the two quark lines in the fundamental basis, the sum over
permutations $\sigma\in S_{n+1}$ in Eq.~\eqref{eq:2qq_color_simple}
includes the permutations of the $n$ gluons, and the
$\overline{q}_2q_2$ pair for $\M_1$ and $\overline{q}_1q_2$ pair for
$\M_2$, respectively, represented by the element $\sigma(n_1+1)$ as
before, and therefore $n_1$ labels the number of $\U(\NC)$ gluons
before the quark-anti-quark pair index.

On top of this, one must also consider external $\U(1)$ gluons by
applying the replacement in Eq.~\eqref{eq:replace} to each of the external gluons in both
terms. However, as in the case of the one quark pair, we only need to
consider up to one external $\U(1)$ gluon to obtain the NLC accurate terms, see Fig.~\ref{fig:gluons}
and the discussion around it. Moreover, following similar arguments,
it can be directly concluded that having both an internal and an
external $\U(1)$ gluon in a dual amplitude cannot result in
colour factors that contribute at NLC. Hence, we only need to consider
up to one external $\U(1)$ gluon for the $\M_1$ term only:
\begin{equation}\label{eq:2qq_replace}
  \M_1 \to \M_1-\frac{1}{\NC}g^{n+2}\sum_{\bar{\sigma}\in \bar{S}_{n+1}}c_1^1(\bar{\sigma})\A_1^1(\bar{\sigma}),
\end{equation}
where
\begin{eqnarray}\label{eq:A11}
\begin{split}
  c_1^1(\bar{\sigma})&=\delta^{i_{q_1}}_{j_{\sigma(1)}}\ldots \delta^{i_{\sigma(n)}}_{j_{q_1}}\delta^{i_{\sigma(n+1)}}_{j_{\sigma(n+1)}}, \\
  \A_1^1(\bar{\sigma})&=\A_{2qq}(q_1,\bar{\sigma}(1),\ldots,\bar{\sigma}(n),\overline{q}_1,\bar{\sigma}(n+1)),
  \end{split}
\end{eqnarray}
and the set of permutations $\bar{S}_{n+1}$ is the subset of the
complete set of permutations $S_{n+1}$ for which $\bar{\sigma}(n+1)$
does not corresponds to the $\overline{q}_2q_2$ pair. Hence, we have
\begin{equation}\label{eq:2qq_color_complete}
  \M_{2qq} =g^{n+2} \left(\sum_{\sigma\in S_{n+1}} c_1(\sigma)\A_1(\sigma)-\frac{1}{\NC}\sum_{\sigma_\in S_{n+1}} c_2(\sigma)\A_2(\sigma)-\frac{1}{\NC}\sum_{\bar{\sigma}\in \bar{S}_{n+1}}c_1^1(\bar{\sigma})\A_1^1(\bar{\sigma})\right),
\end{equation}
where we have neglected terms that do not contribute up to NLC
accuracy.

We obtain then the colour-summed squared amplitude, in block matrix
notation,
\begin{multline}\label{eq:2qq_color_matrix}
  |\M_{2qq}|^2 = (g^2)^{n+2}\sum_{\sigma_k,\sigma_l}
  \begin{pmatrix}
    \A_1(\sigma_k) & \A_2(\sigma_k) & \A_1^1(\bar{\sigma}_k)
  \end{pmatrix}  \\
  \begin{pmatrix}
    c_1(\sigma_k)c_1(\sigma_l)^{\dagger}  & -c_1(\sigma_k)c_2(\sigma_l)^{\dagger}\!/\NC& -c_1(\sigma_k)c_1^1(\bar{\sigma}_l)^{\dagger}\!/\NC \\
    -c_2(\sigma_k)c_1(\sigma_l)^{\dagger}\!/\NC  & c_2(\sigma_k)c_2(\sigma_l)^{\dagger}\!/\NC^2& c_2(\sigma_k)c_1^1(\bar{\sigma}_l)^{\dagger}\!/\NC^2\\
    -c_1^1(\bar{\sigma}_k)c_1(\sigma_l)^{\dagger}\!/\NC & c_1^1(\bar{\sigma}_k)c_2(\sigma_l)^{\dagger}\!/\NC^2& c_1^1(\bar{\sigma}_k)c_1^1(\bar{\sigma}_l)^{\dagger}\!/\NC^2
  \end{pmatrix}
  \begin{pmatrix}
    \A_1(\sigma_l)^* \\
    \A_2(\sigma_l)^* \\
    \A_1^1(\bar{\sigma}_l)^*
  \end{pmatrix},
\end{multline}
where, with some abuse of notation, we have written it with a double
sum over $\sigma_k$ and $\sigma_l$, with the corresponding subpermutations
$\bar{\sigma}_k$ and $\bar{\sigma}_l$ understood.

\subsubsection*{Leading colour}
The leading-colour contribution comes from the similar structure as
for the one-quark-line case, i.e.~the upper-left block of
Eq.~\eqref{eq:2qq_color_matrix}
\begin{multline}\label{eq:LC1}
  \A_{2qq}(q_1,\sigma_k(1),\ldots ,\sigma_k(n_1),\overline{q}_2q_2,\sigma_k(n_1+2),\ldots ,\sigma_k(n),\overline{q}_1) \times \\
  \left(\A_{2qq}(q_1,\sigma_l(1),\ldots ,\sigma_l(n_1),\overline{q}_2q_2,\sigma_l(n_1+2),\ldots ,\sigma_l(n),\overline{q}_1)\right)^*,
\end{multline}
with $\sigma_k=\sigma_l$. This comes with a colour factor equal to
\begin{equation}\label{eq:LC2}
  \delta^{i_{q_1}}_{j_{\sigma_k(1)}}\delta^{i_{\sigma_k(1)}}_{j_{\sigma_k(2)}}\ldots \delta^{i_{\sigma_k(n+1)}}_{j_{q_1}} 
  \left(\delta^{i_{q_1}}_{j_{\sigma_l(1)}}\delta^{i_{\sigma_l(1)}}_{j_{\sigma_l(2)}}\ldots \delta^{i_{\sigma_l(n+1)}}_{j_{q_1}}\right)^{\dagger}  = \NC^{n+2}.
\end{equation}

\subsubsection*{Next-to-leading colour}
Next-to-leading colour contributes as $\pm\NC^{n}$ (with an exception
mentioned below), where the $\pm$ corresponds to the sign of the terms
in the colour matrix of Eq.~\eqref{eq:2qq_color_matrix}. We consider
each of the blocks in that colour matrix separately.
\begin{itemize}
\item The upper-left block has a similar structure as to the one quark
  pair case, albeit with an additional $\overline{q}_2q_2$ pair in the
  permutations. This can give a NLC contribution if and only if
  \begin{equation}
\sigma_R \sigma_{Q_1} \sigma_S \sigma_{Q_2} \sigma_P \times \left(\sigma_R \sigma_{Q_2} \sigma_S \sigma_{Q_1} \sigma_P\right)^{\dagger}.
  \end{equation}
\item The upper-middle and middle-left blocks are interference
  contributions in which the quark order in the dual amplitude and
  conjugate amplitude is different. Similarly to the fundamental basis
  (c.f.~upper-right and lower-left blocks of Eq.~\eqref{eq:2qq_fund})
  these terms can contribute at NLC. Let
  \begin{align}
    \sigma_A=\sigma(1),\ldots,\sigma(n_1) \quad\textrm{and}\quad
    \sigma_B=\sigma(n_1+2),\ldots,\sigma(n+1),
  \end{align}
  that is, the gluon subpermutations of $\sigma_k$ (i.e.~the dual
  amplitude) before and after the quark-pair index, respectively. In
  order for the $\sigma_l$ permutation to give a NLC contribution, it
  must be that
  \begin{align}\label{eq:2qq_color_split}
    \sigma_C=&\sigma_A^1\sigma_B^2 \quad\textrm{and}\quad \sigma_D=\sigma_B^1\sigma_A^2,
  \end{align}
  where $\sigma_C$ and $\sigma_D$ are, respectively, the
  subpermutations of the gluons before and after the quark-pair index
  of the $\sigma_l$ permutation, i.e.~of the conjugate dual
  amplitude. The superscripts 1 (2) denote the first (second)
  subpermutation after splitting the $\sigma_{A,B}$ in two in any
  possible way, including one of the parts being the empty
  permutation.

\item The lower-right block is the contribution in which both the
  amplitude and the conjugate amplitude have an external $\U(1)$
  gluon. It contributes at NLC with entries $-(n+1)\NC^{n}$ in the
  colour matrix if and only if $\bar{\sigma}_k=\bar{\sigma}_l$.
  
\item The lower-left block contains interference contributions in
  which only the amplitude has an external $\U(1)$ gluon. The quark
  order is identical for the amplitude and the conjugate amplitude,
  and therefore this is very similar to the one-quark-pair case. These
  terms are canceled in the colour matrix, owing to the reduction of
  the same type as in Eq.~\eqref{eq:sum_U1} but for the two-quark-pair
  correspondence.
  
\item The center block comes with an explicit $1/\NC^2$ suppression,
  and can therefore only contribute to NLC with $\NC^{n}$ if the
  strings of Kronecker deltas result in $n+2$ colour loops. This
  happens only if $\sigma_k=\sigma_l$.

\item Finally, the middle-right and lower-middle blocks only contribute beyond
  NLC.
\end{itemize}
This completes all the blocks of Eq.~\eqref{eq:2qq_color_matrix}.

\subsubsection*{Proof of the NLC terms}
Let us start by discussing the upper-middle and middle-left
blocks. That is, for which the anti-quark ordering is different in the
dual amplitudes and conjugate amplitudes. The colour factors already
come with an explicit $1/\NC$ suppression. Furthermore, due to the
different ordering of the anti-quark indices, the corresponding
Kronecker deltas cannot form $n+2$ colour loops: there will be at
least one loop made from four Kronecker deltas resulting in a maximum
of $n+1$ loops. That introduces another reduction of at least $1/\NC$
in comparison to LC. Therefore, in order for these terms to contribute
at NLC, all the other $2n$ Kronecker deltas must yield $n+1$ colour
loops. This can only happen if the subpermutation that corresponds to
these Kronecker deltas is identical in the dual amplitude and
conjugate amplitude. In other words, apart from a single split and
interchange, the $\sigma_C$ and $\sigma_D$ subpermutations must be
identical to $\sigma_A$ and $\sigma_B$, respectively, see
eq.~\eqref{eq:2qq_color_split}.

Similarly to the one-quark-line case the lower-left and the
upper-right blocks can be reduced and absorbed in the lower-left
block.  For the lower-left block, the interference contributes at NLC
if and only if the colour order of the $n-1$ $\U(\NC)$~gluons and
$\overline{q}_2q_2$ pair in the amplitude is identical to the order of
these gluons and $\overline{q}_2q_2$ pair in the conjugate amplitude,
while the $\U(1)$ gluon might be inserted at any location as a
$\U(\NC)$~gluon in the order of the conjugate amplitude. This yields a
total of $n+1$ possible conjugate amplitudes. In the same manner as
was done in the one-quark pair case, these interferences can however
be reduced, since in a single row with a dual amplitude of $\A_1^1$
type, all interferences between amplitudes of type $\A_1$ which yield
a NLC is
  \begin{eqnarray}
  \begin{split}
-\NC^{n} \A_{2qq}(q_1,\bar{\sigma}_k(1),\ldots,\bar{\sigma}_k(n),&\overline{q}_1,\bar{\sigma}_k(n+1))  \times \\ 
 &\left[ (\A_{2qq}(q_1,\bar{\sigma}_k(n+1),\sigma_k(1),\ldots , \bar{\sigma}_k(n),\overline{q}_1))^* + \right.\\
 &\left. (\A_{2qq}(q_1,\sigma_k(1),\bar{\sigma}_k(n+1),\ldots ,\ldots,\bar{\sigma}_k(n),\overline{q}_1))^* +  \right. \\
 &\left. \ldots \right. \\
 &\left. (\A_{2qq}(q_1,\sigma_k(1),\ldots ,\ldots,\bar{\sigma}_k(n),\bar{\sigma}_k(n+1),\overline{q}_1))^* \right]
 \end{split}
  \end{eqnarray}
where in the square parenthesis the sum is precisely that of
$(\A_{2qq}(q_1,\bar{\sigma}_k(1),\ldots,\bar{\sigma}_k(n),\overline{q}_1,\bar{\sigma}_k(n+1)))^*$. Therefore,
the sum of these contributions cancel the interference in the
lower-right block.
  
A similar analysis is done for the upper-right block, with a dual
amplitude of $\A_1$ and conjugate amplitudes of $\A_1^1$
types. Exactly as in the one-quark pair case, these can be reduced to
be included in the diagonal elements in the lower-right block, with
colour factor $-(n+1)\NC^{n}$. This completes the proof.

\subsection{Two same-flavour  $q\overline{q}$ pairs and $n$ gluons}
For the same-flavour two-quark-line case in the colour-flow
decomposition the arguments follow closely those of the fundamental
basis. However, now we must also include external $\U(1)$ gluon
emission in the quark ordered amplitudes. For the NLC contributions,
it is sufficient to consider only single $\U(1)$ emission. Using the
same notation as in the distinct-flavour case in the previous section, we have
\begin{align}
\M_1 =&g^{n+2} \sum_{\sigma\in S_{n+1}} c_1(\sigma) \A_1(\sigma) - \frac{1}{\NC}g^{n+2} \sum_{\bar{\sigma} \in \bar{S}_{n+1}} c_1^1(\bar{\sigma}) \A_1^1(\bar{\sigma}), \\
\M_2 =& g^{n+2} \sum_{\sigma \in S_{n+1}} c_2(\sigma) \A_2(\sigma) - \frac{1}{\NC}g^{n+2} \sum_{\bar{\sigma} \in \bar{S}_{n+1}} c_2^1(\bar{\sigma}) \A_2^1(\bar{\sigma}).
\end{align}
Symmetrising the distinct-flavour results, we can write the amplitude as 
\begin{multline}\label{eq:2qq_SF_col_amp_types}
\M_{2qq} = \left( 1+\frac{1}{\NC} \right) g^{n+2} \bigg(\sum_{\sigma\in S_{n+1}} c_1(\sigma) \A_1(\sigma)-\frac{1}{\NC}\sum_{\bar{\sigma} \in \bar{S}_{n+1}} c_1^1(\bar{\sigma}) \A_1^1(\bar{\sigma}) -  \\
 \sum_{\sigma \in S_{n+1}} c_2(\sigma) \A_2(\sigma) + \frac{1}{\NC}\sum_{\bar{\sigma}\in \bar{S}_{n+1}} c_2^1(\bar{\sigma}) \A_2^1(\bar{\sigma}) \bigg).
\end{multline}
This yields a colour-summed squared amplitude, in block matrix
notation, equal to
\begin{eqnarray}
\begin{split}
& |\M_{2qq}|^2 = (g^2)^{n+2} \left(1+\frac{1}{\NC} \right)^2
& \sum_{\sigma_k,\sigma_l} 
\begin{pmatrix}
\A_1(\sigma_k)& \A_2(\sigma_k) & \A_1^1(\bar{\sigma}_k) & \A_2^1(\bar{\sigma}_k)
\end{pmatrix}
\mathbb{C}
\begin{pmatrix}
\A_1(\sigma_l)^* \\
\A_2(\sigma_l)^* \\
\A_1^1(\bar{\sigma}_l)^* \\
\A_2^1(\bar{\sigma}_l)^*
\end{pmatrix}
\end{split}
\end{eqnarray}
with the colour matrix
\begin{eqnarray}\label{eq:2qqSF_colmat}
\mathbb{C}= \begin{pmatrix}
 c_1(\sigma_k) c_1(\sigma_l)^{\dagger}  &  -c_1(\sigma_k)c_2(\sigma_l)^{\dagger}  & -c_1(\sigma_k) c_1^1(\bar{\sigma}_l)^{\dagger}\!/\NC & c_1(\sigma_k) c_2^1(\bar{\sigma}_{l})^{\dagger}\!/\NC \\
-c_2(\sigma_k) c_1(\sigma_l)^{\dagger}  &   c_2(\sigma_k)c_2(\sigma_l)^{\dagger}  & 
-c_2(\sigma_k) c_1^1(\bar{\sigma}_l)^{\dagger}\!/\NC & 
-c_2(\sigma_k) c_2^1(\bar{\sigma}_l)^{\dagger}\!/\NC \\
-c_1^1(\bar{\sigma}_k) c_1(\sigma_l)^{\dagger}\!/\NC &
  c_1^1(\bar{\sigma}_k) c_2(\sigma_l)^{\dagger}\!/\NC   &
   c_1^1(\bar{\sigma}_k) c_1^1(\bar{\sigma}_l)^{\dagger}\!/\NC^2 & -c_1^1(\bar{\sigma}_k) c_2^1(\bar{\sigma}_l)^{\dagger}\!/\NC^2 \\
 c_2^1(\bar{\sigma}_k) c_1(\sigma_l)^{\dagger}\!/\NC &  
 -c_2^1(\bar{\sigma}_k) c_2(\sigma_l)^{\dagger}\!/\NC   & 
 -c_2^1(\bar{\sigma}_k) c_1^1(\bar{\sigma}_l)^{\dagger}\!/\NC^2 &c_2^1(\bar{\sigma}_k) c_2^1(\bar{\sigma}_l)^{\dagger}\!/\NC^2
\end{pmatrix}.
\end{eqnarray}
We notice that in this case, the colour factors are no longer strictly
monomials in $\NC$, but rather a monomial times $(1+1/\NC)^2$. As
in the fundamental decomposition, we consider contributions to be next-to-leading colour if they
are up to $1/\NC^2$-suppressed from the leading-colour one, hence they
are of order $\mathcal{O}(\NC^{n+1})$ and/or $\mathcal{O}(\NC^{n})$.

\subsubsection*{Leading colour}
Only the first and second diagonal blocks of the colour matrix,
Eq.~\eqref{eq:2qqSF_colmat}, contribute at LC, and only if
$\sigma_k=\sigma_l$.

\subsubsection*{Next-to-leading colour}
Comparing the colour matrix in Eq.~\eqref{eq:2qqSF_colmat} with the
colour matrix obtained in the distinct-flavour case in Eq.~\eqref{eq:2qq_color_matrix}, the only really
new terms that appear are those with colour factors:
$c_2^1(\bar{\sigma}_k)
c_1(\bar{\sigma}_l)^{\dagger}\!/\NC$ and
$c_2^1(\bar{\sigma}_k)
c_1^1(\bar{\sigma}_l)^{\dagger}\!/\NC^2$ , and the corresponding conjugate cases.
However, none of these new colour factors yield a NLC element. The remaining blocks in the last row of the colour matrix are $-c_2^1(\bar{\sigma}_k)
c_2(\bar{\sigma}_l)^{\dagger}\!/\NC$ and the diagonal $c_2^1(\bar{\sigma}_k)
c_2^1(\bar{\sigma}_l)^{\dagger}\!/\NC^2$. The sum of the former contributions can be combined to cancel the contribution from the latter type, as was performed in the distinct flavour case. Finally, the conjugate terms $-c_2(\bar{\sigma}_k)
c_2^1(\bar{\sigma}_l)^{\dagger}\!/\NC$ are re-located to the diagonal terms of the $-c_2^1(\bar{\sigma}_k)
c_2^1(\bar{\sigma}_l)^{\dagger}\!/\NC^2$ block with the entry $-(n+1)\NC^{n}$.  

Furthermore, for the other elements
in the colour matrix, we have different explicit $1/\NC$ suppression
w.r.t.~the colour matrix of the distinct-flavour case. However, these
differences are at most only one power of $1/\NC$, while different
contractions of the strings of deltas comes with two powers of
$1/\NC$. Using the definition of NLC to contain terms of $\mathcal{O}(\NC^{n+1})$ and $\mathcal{O}(\NC^{n})$, but not $\mathcal{O}(\NC^{n-1})$, there are no additional pieces arising as compared to the distinct-flavour case and the arguments given in there apply here without the need for any modification.

\section{Results}\label{sec:results}
In this section we present the main result of this work by showing the
number of dual amplitudes which need to be computed in order to
calculate scattering probabilities at NLC accuracy. That is, the
number of terms in the double sum of Eq.~\eqref{eq:colourdecomp} that
needs to be considered to get a result that is correct up to NLC
accuracy at tree-level.

\subsection{Phase-space symmetrisation}\label{sec:symmetry}
Since final state identical particles are indistinguishable,
interchanging the momenta of these particles in the numerical
phase-space integration must yield identical matrix elements. This
fact can be used to reduce the number of terms in the sum of
Eq.~\eqref{eq:colourdecomp} that needs to be computed, irrespective of
the expansion in colour. For example, for $gg\to (n-2)\, g$
scattering, there is a $1/(n-2)!$ symmetry factor due to the $n-2$
identical final state gluons. Therefore, if the phase-space
generation in the numerical integration is symmetric under interchange
of any two of these gluons, the sum over $k$ (or $l$) can be reduced
by the $(n-2)!$ factor, since, upon integration, these rows (or
columns) will yield identical results to the cross section and
differential distributions~\cite{Badger:2012pg}.

The number of identical final state gluons ($n_g$), quarks ($n_{q}$)
and anti-quarks ($n_{\bar{q}}$) depend on the scattering process under
consideration.  In Tab.~\ref{tab:symmetry_fac} we list these values
for the different types of QCD-particle-initiated processes, together
with the total number of dual amplitudes (the number of rows/columns
of the colour matrix). The number of non-zero elements in the colour matrix in the fundamental and colour-flow decompositions, including the phase-space symmetrisation, is discussed in the following.

\renewcommand*{\arraystretch}{1.2}
 \begin{table}[htb!]
 \center
 \begin{tabular}{l  r r r r r r }
 \toprule
\multicolumn{1}{l}{Process} & Total  (fund.)  & Total (colour-flow) &    $n_g$   &     $n_{q}$  & $n_{\bar{q}}$ \\
\midrule
$gg \rightarrow (n-2)\,g$ &    $(n-1)!$ & $(n-1)!$ &   $n-2$  &    0 & 0 \\
$gg \rightarrow q\overline{q}+(n-2)\,g$  &  $n!$ &  $n!\sum_{r=0}^n \frac{1}{r!}$ & $n-2$  & 0 & 0 \\
$gg \rightarrow q\overline{q}Q\overline{Q}+(n-2)\,g$  & $2(n+1)!$ &   $2n! \sum_{r=0}^n \frac{1}{r!}(n+1-r)$ &   $n-2$ & 0 & 0 \\
$gg \rightarrow q\overline{q}q\overline{q}+(n-2)\,g$  & $2(n+1)!$ &    $2n! \sum_{r=0}^n \frac{1}{r!}(n+1-r)$ &$n-2$& 2 & 2\\
[0.3cm]
$qg/\overline{q}g \rightarrow q/\overline{q}+(n-1)\,g$ & $n!$    &$n!\sum_{r=0}^n \frac{1}{r!}$ & $n-1$ & 0 & 0 \\
$qg/\overline{q}g \rightarrow q/\overline{q} \ Q \overline{Q}+(n-1)\,g$ & $2(n+1)!$ &  $2n! \sum_{r=0}^n \frac{1}{r!}(n+1-r)$   &  $n-1$ & 0 & 0\\
$qg\rightarrow qq \overline{q}+(n-1)\,g$& $2(n+1)!$ & $2n! \sum_{r=0}^n \frac{1}{r!}(n+1-r)$ & $n-1$  & 2 & 0 \\
$\overline{q}g \rightarrow q\overline{q}\overline{q}+(n-1)\,g$& $2(n+1)!$ & $2n! \sum_{r=0}^n \frac{1}{r!}(n+1-r)$ & $n-1$  & 0 & 2 \\
 [0.3cm]
$q\overline{q}  \rightarrow n\,g$ & $n!$  & $n!\sum_{r=0}^n \frac{1}{r!}$ &  $n$   & 0 & 0 \\
$q\overline{q}  \rightarrow Q \overline{Q}/ q \overline{q}+n\,g$ & $2(n+1)!$ & $2n! \sum_{r=0}^n \frac{1}{r!}(n+1-r)$ & $n$   & 0 & 0\\
$qQ/q\overline{Q} \rightarrow  qQ/q\overline{Q} + n\,g$ & $2(n+1)!$ & $2n! \sum_{r=0}^n \frac{1}{r!}(n+1-r)$ &$n$  & 0 & 0\\
$qq \rightarrow  qq + n\,g$  & $2(n+1)!$ & $2n! \sum_{r=0}^n \frac{1}{r!}(n+1-r)$ & $n$  & 2 & 0 \\
$\overline{q}\overline{q} \rightarrow  \overline{q}\overline{q} + n\,g$  & $2(n+1)!$ & $2n! \sum_{r=0}^n \frac{1}{r!}(n+1-r)$ & $n$  & 0 & 2 \\
 \bottomrule
\end{tabular}
 \caption{The total number of rows in the colour matrix in the
   fundamental (second column) and colour-flow (third column)
   decomposition for various multi-parton processes at hadron
   colliders. The number of identical final state gluons ($n_g$) and
   final state quarks ($n_{q}$) and anti-quarks ($n_{\bar{q}}$) are
   given in the last three columns.}
\label{tab:symmetry_fac}
\end{table}

\subsubsection*{Fundamental decomposition}
The total number of independent rows in the colour matrix after
phase-space symmetrisation for various processes is given by
\begin{eqnarray}\label{eq:sym_fac_F}
\frac{N}{n_g!\,n_{q}!\,n_{\bar{q}}!},
\end{eqnarray}
with $N$ the total number of rows, listed in the second column of
Tab.~\ref{tab:symmetry_fac}, and the number of identical final state
gluons $n_g$ and the symmetry factor from identical final state quarks
$n_{q}$ and anti-quarks $n_{\bar{q}}$, listed in the last three
columns of Tab.~\ref{tab:symmetry_fac}.

\subsubsection*{Colour-flow decomposition}
As shown in Tab.~\ref{tab:symmetry_fac}, the size of the colour matrix
for all-gluon processes is the same in the colour-flow decomposition
as in the fundamental decomposition, and the reduction due to the
phase-space symmetrisation is the same as in
Eq.~\eqref{eq:sym_fac_F}. For amplitudes with external quark lines,
however, the case is more subtle. For processes with one quark line
plus $n$ gluons the number of amplitudes differs due to the
possibility for external $\U(1)$ gluons, which are considered
as different dual amplitudes than their corresponding $\U(\NC)$
counterparts. The total number of dual amplitudes in this case is
\begin{eqnarray}\label{eq:total_amp1qq}
n!\sum_{r=0}^n \frac{1}{r!}
\end{eqnarray}
where the sum is over the number of external $\U(1)$ gluons in the
amplitude. The factors $1/r!$ take into consideration that the $\U(1)$
gluons, being colourless particles, do not take part in the
colour-ordering and therefore their permutations do not yield
different dual amplitudes. We can rewrite Eq.~\eqref{eq:total_amp1qq} as
\begin{equation}\label{eq:total_amp1qq_v2}
  \sum_{r=0}^{n}\, \sum_{k=k_{\textrm{min}}}^{k_{\textrm{max}}}
  \binom{n_g}{k} \binom{n-n_g}{r-k}(n-r)!
\end{equation}
where $k$ is the number of $\U(1)$ gluons among the $n_g$ final state
gluons, and the sum over $k$ is therefore between
$k_{\textrm{min}}=\text{max}(0,r-(n-n_g))$ and
$k_{\textrm{max}}=\text{min}(r,n_g)$.

The reduction due to the symmetrisation of the phase-space applies to
all final state gluons, irrespective if they are $\U(1)$ and $\U(\NC)$
gluons. However, care must be taken since permutations among the
$\U(1)$ gluons yield identical dual amplitudes, which is already taken
into account in
Eqs.~\eqref{eq:total_amp1qq}~and~\eqref{eq:total_amp1qq_v2}. Hence,
the reduction is only equal to $k!/n_g!$.  Therefore, the total number
of independent rows for processes with one quark pair is
\begin{eqnarray}\label{eq:col_tot1qq}
  \sum_{r=0}^{n}\, \sum_{k=k_{\textrm{min}}}^{k_{\textrm{max}}}
  \binom{n_g}{k} \binom{n-n_g}{r-k}(n-r)!\frac{k!}{n_g!}.
\end{eqnarray}
At NLC we only have zero or one external $\U(1)$ gluon, i.e.~$r=0,1$,
and the sum in Eq.~\eqref{eq:col_tot1qq} simplifies considerably to $2 n! /
n_g!$.

In the two-quark-line amplitudes, one similarly has to take into
consideration the external $\U(1)$ gluons but also the partition of
gluons among the colour lines. The total number of dual amplitudes is
\begin{eqnarray}
2n! \sum_{r=0}^n \frac{1}{r!}(n+1-r),
\end{eqnarray}
with the factor $(n+1-r)$ considering the partition of the gluons on
the two colour lines. Comparing this expression to
Eq.~\eqref{eq:total_amp1qq}, results in
\begin{eqnarray}\label{eq:col_tot2qq}
  \frac{2}{n_{q}!\,n_{\bar{q}}!} \sum_{r=0}^{n}\, \sum_{k=k_{\textrm{min}}}^{k_{\textrm{max}}}
  \binom{n_g}{k} \binom{n-n_g}{r-k}(n-r)!\frac{k!}{n_g!}(n+1-r),
\end{eqnarray}
independent rows in the colour matrix when taking the phase-space
symmetrisation into account for gluons and (anti-)quarks. Note that
there is one subtlety regarding the (anti-)quark symmetry factors. For
the two-quark-line same-flavour case in the $gg$ initiated process, we
have a quark symmetry factor of $n_{q}!\,n_{\bar{q}}!=4$. We note
however that not all contributions to the dual amplitudes recover this
symmetry. In particular, the case where all gluons are $\U(1)$ gluons,
the interchange of the quark labels $q_1 \leftrightarrow q_2$,
$\overline{q}_1 \leftrightarrow \overline{q}_2$ recovers the same
amplitude, since the two colour lines are now identical with no
$\U(\NC)$ gluons attached. Hence, to obtain the number of independent
rows in this case, one must divide by the symmetry factor
$n_{q}!\,n_{\bar{q}}!=4$ only the contributions with $r=0,\ldots,n-1$
and divide by 2 the $r=n$ contribution.

\subsection{Number of non-zero elements in the colour matrix}
We start this section by presenting the number of dual conjugate amplitudes
that need to be considered for the computation of a single dual amplitude
to yield all contributions up to NLC accuracy. In other words this
corresponds to computing the number of non-zero elements in a single
row (or column) in the colour matrix defined in
Eq.~\eqref{eq:colourdecomp} up to NLC accuracy. This combined with the
number of independent rows due to phase-space symmetrisation, as
discussed in the previous section, yields the total number of terms
that need to be considered.

\renewcommand*{\arraystretch}{1.2}
 \begin{table}[htb!]
 \center
 \begin{tabular}{ r r @{~} l r @{~} l r @{~} l }
 \toprule
 \multicolumn{7}{l}{all-gluon}\\
 $n$   & \multicolumn{2}{l}{Fundamental} & \multicolumn{2}{l}{Colour-flow} & \multicolumn{2}{c}{Adjoint} \\
 \midrule
  4  &    6 &          (6) &    6 &          (6) &     2 &     (2)\\ 
  5  &   11 &         (24) &   16 &         (24) &     5 &     (6)\\
  6  &   24 &        (120) &   36 &        (120) &    18 &    (24)\\
  7  &   50 &        (720) &   71 &        (720) &    93 &   (120)\\
  8  &   95 &       (5040) &  127 &       (5040) &   583 &   (720)\\
  9  &  166 &      (40320) &  211 &      (40320) &  4162 &  (5040)\\
 10  &  271 &     (362880) &  331 &     (362880) & 31649 & (40320)\\
 11  &  419 &    (3628800) &  496 &    (3628800) &     - & \\
 12  &  620 &   (39916800) &  716 &   (39916800) &     - & \\
 13  &  885 &  (479001600) & 1002 &  (479001600) &     - & \\
 14  & 1226 & (6227020800) & 1366 & (6227020800) &     - & \\
 \bottomrule
\end{tabular}
 \caption{Number of non-zero elements in a single row of the colour
   matrix for all-gluon matrix elements up to NLC accuracy,
   $\mathcal{O}(N_c^{n-2})$, in the fundamental, colour-flow and
   adjoint bases (the latter only up to $n=10$ for computational time
   reasons). The number between brackets is the total number of
   elements, i.e.~the number needed when full-colour is desired.}
\label{tab:gluons}
\end{table}
In Tab.~\ref{tab:gluons} we present the number of non-zero terms in a
single row of the colour matrix for the all-gluon matrix elements. The
first column lists the number of gluons in the process. In the second
column the number of non-zero terms at NLC are presented using the
fundamental basis, while in the third column the same is presented for
the colour-flow basis. For completeness, in the final column we also
show the number of terms needed in the adjoint representation to
obtain NLC accuracy (for numerical reasons, only up to
$n=10$). For the three cases, the number in brackets is the
total number of elements in a single row, i.e., the number that needs
to be computed to obtain full-colour accuracy. It is interesting to
note that the scaling with $n$, the number of gluons, is reduced from
factorial (the numbers in brackets) to polynomial, $n^4$, in both the
fundamental and colour-flow bases when considering the matrix elements
only up to NLC accuracy. Somewhat surprisingly, in the adjoint basis
there is no such reduction and the number of terms still scales
factorially. This means that even though this basis is optimal when
computing the full-colour matrix elements, since there are only
$(n-2)!$ terms, as compared to the $(n-1)!$ in the fundamental and
colour-flow bases, this is no longer true at NLC accuracy with
multiple gluons. Comparing the fundamental and colour-flow bases, it
can be seen that the fundamental basis is slightly more efficient:
there are fewer non-zero elements in a single row of the colour matrix
to compute. As explained in
Secs.~\ref{sec:gluon}~and~\ref{sec:gluon_CF}, the reason is the list
of exceptions to the general rule that interchanging a single
subpermutation of a (string of) generators gives a NLC contribution
when using the fundamental basis. A similar rule applies to the
colour-flow basis, but without the exceptions, resulting in the need
to compute slightly more terms.

\renewcommand*{\arraystretch}{1.2}
 \begin{table}[htb!]
 \center
 \begin{tabular}{ r r @{~} l @{\qquad} r @{~} l r @{~} l }
 \toprule
\multicolumn{5}{l}{$q\overline{q} + n\,g$} \\
 \multirow{2}{*}{$n$}  & \multicolumn{2}{l}{\multirow{2}{*}{Fundamental}} & \multicolumn{3}{c} {\qquad\qquad\qquad Colour-flow} \\
  &     \multicolumn{2}{l}{}     &       \multicolumn{2}{l}{no external $\U(1)$} & \multicolumn{2}{l}{one external $\U(1)$} \\
 \midrule
 2  &   2 &          (2) &    2 & (5)          &  1 & (5)          \\ 
 3  &   4 &          (6) &    6 & (16)         &  1 & (16)         \\ 
 4  &  10 &         (24) &   16 & (65)         &  1 & (65)         \\ 
 5  &  24 &        (120) &   36 & (326)        &  1 & (326)        \\
 6  &  51 &        (720) &   71 & (1957)       &  1 & (1957)       \\
 7  &  97 &       (5040) &  127 & (13700)      &  1 & (13700)      \\
 8  & 169 &      (40320) &  211 & (109601)     &  1 & (109601)     \\
 9  & 275 &     (362880) &  331 & (986410)     & 1 & (986410)     \\
10  & 424 &    (3628800) &  496 & (9864101)    & 1 & (9864101)    \\
11  & 626 &   (39916800) &  716 & (108505112)  & 1 & (108505112)  \\
12  & 892 &  (479001600) & 1002 & (1302061345) & 1 & (1302061345) \\
\bottomrule
\end{tabular}
 \caption{Number of non-zero elements in a single row of the colour
   matrix for $q\overline{q} + n\,g$ matrix elements up to NLC
   accuracy, $\mathcal{O}(N_c^{n-1})$, in the fundamental and
   colour-flow bases. The number between brackets is the total number
   of elements, i.e.~the number needed when full-colour is
   desired. For the colour-flow basis, we distinguish the cases in
   which the row in the colour matrix corresponds to an amplitude with
   no external $\U(1)$ gluon, or with one external $\U(1)$
   gluon.}
\label{tab:1qq}
\end{table}
In Tab.~\ref{tab:1qq} we present the number of dual conjugate
amplitudes that need to be computed for a single dual amplitude to
obtain NLC accuracy for $q\overline{q} + n\,g$ matrix elements. In the
first column, there is the number of gluons $n$ present in the
scattering process. The second column lists the number of non-zero
terms in a single row of the colour matrix when using the fundamental
basis, with the total number of terms in the row in brackets. For the
colour-flow basis, we need to distinguish if the corresponding dual
amplitude contains an external $\U(1)$ gluon or not, since the number of
conjugate amplitudes that need to be considered is different in the two
cases. In the fundamental basis, there are a total number of $n!$
different amplitudes. Hence, for a full colour computation all these
contribute to a single row of the colour matrix, and is given in
brackets in the second column of the table. The number of conjugate amplitudes is reduced to a
polynomial $n^4$ scaling when reducing the colour accuracy to NLC,
just as in the case of all-gluon matrix elements. In the colour-flow
basis, if the dual amplitude does not contain any $\U(1)$ gluons, the
NLC terms are produced if the colour ordering of the conjugate
amplitude has the form given by Eq.~\eqref{eq:1qq_pos1_CF}. This also
reduces the number of terms to a polynomial $n^4$ scaling. On the
other hand, if the amplitude contains a (single) $\U(1)$ gluon, using the reduction presented in Sec. \ref{sec:color-flow-1quark}, all interferences are moved to the diagonal of these rows, leaving only one conjugate amplitude (column) needed for each row of this type. Note that in the full-colour approach,
the number of terms that need be considered scales worse than in the
fundamental basis, since not only all permutation of colour ordering
for the gluons need be considered (which results in the $n!$ number of
terms of the fundamental basis), but also all the replacements of
$\SU(N_c)$~gluons by $\U(1)$~gluons must be taken into account.

\begingroup
\renewcommand*{\arraystretch}{1.2}
 \begin{table}[htb!]
 \center
 \begin{tabular}{  l r@{\,\big|\,}l r@{\,\big|\,}l r@{\,\big|\,}l r@{\,\big|\,}l r@{\,\big|\,}l r@{\,\big|\,}l    @{\qquad} l }
 \toprule
 \multicolumn{4}{l}{$q\overline{q}\,Q\overline{Q}\,+\,n\,g$} & \multicolumn{10}{r}{Fundamental:\qquad $\A_1$ \big| $\A_2$ types} \\
 \multirow{2}{*}{$n$} & \multicolumn{12}{c} {$\min(n_1,n-n_1)$} & \\
& \multicolumn{2}{c}{$0$} & \multicolumn{2}{c}{$1$} & \multicolumn{2}{c}{$2$} & \multicolumn{2}{c}{$3$} & \multicolumn{2}{c}{$4$} & \multicolumn{2}{c}{$5$} &  \\
 \midrule
 0 & 2 & 2 & \multicolumn{2}{c}{ } & \multicolumn{2}{c}{ } & \multicolumn{2}{c}{ } & \multicolumn{2}{c}{ } & \multicolumn{2}{c}{ } & (2) \\ 
 1 & \textbf{3} & \textbf{3} & \multicolumn{2}{c}{ } & \multicolumn{2}{c}{ } & \multicolumn{2}{c}{ } & \multicolumn{2}{c}{ } & \multicolumn{2}{c}{ } & (4) \\ 
 2 & \textbf{7} & \textbf{4} & 6 & 5 & \multicolumn{2}{c}{ } & \multicolumn{2}{c}{ } & \multicolumn{2}{c}{ } & \multicolumn{2}{c}{ } & (12) \\ 
 3 & \textbf{15} & \textbf{5} & \textbf{15} & \textbf{7} & \multicolumn{2}{c}{ } & \multicolumn{2}{c}{ } & \multicolumn{2}{c}{ } & \multicolumn{2}{c}{ } & (48) \\ 
 4 & \textbf{31} & \textbf{6} & \textbf{32} & \textbf{9} & 33 & 10 & \multicolumn{2}{c}{ } & \multicolumn{2}{c}{ } & \multicolumn{2}{c}{ } & (240) \\
 5 & \textbf{60} & \textbf{7} & \textbf{62} & \textbf{11} & \textbf{64} & \textbf{13} & \multicolumn{2}{c}{ } & \multicolumn{2}{c}{ } & \multicolumn{2}{c}{ } & (1440) \\ 
 6 & \textbf{108} & \textbf{8} & \textbf{111} & \textbf{13} & \textbf{114} & \textbf{16} & 115 & 17 & \multicolumn{2}{c}{ } & \multicolumn{2}{c}{ } & (10080) \\ 
 7 & \textbf{182} & \textbf{9} & \textbf{186} & \textbf{15} & \textbf{190}& \textbf{19} & \textbf{192} & \textbf{21} & \multicolumn{2}{c}{ } & \multicolumn{2}{c}{ } & (80640) \\
 8 & \textbf{290} & \textbf{10} & \textbf{295} & \textbf{17} & \textbf{300} & \textbf{22} & \textbf{303} & \textbf{25} & 304 & 26 & \multicolumn{2}{c}{ } & (725760) \\
 9 & \textbf{441} & \textbf{11} & \textbf{447} & \textbf{19} & \textbf{453} & \textbf{25} & \textbf{457} & \textbf{29} & \textbf{459} & \textbf{31} & \multicolumn{2}{c}{ } & (7257600) \\
 10 & \textbf{645} & \textbf{12} & \textbf{652} & \textbf{21} & \textbf{659} & \textbf{28} & \textbf{664} & \textbf{33} & \textbf{667} & \textbf{36} & 668 & 37 & (79833600) \\
 \bottomrule
\end{tabular}
 \caption{Number of non-zero elements in a single row of the colour
   matrix for $q\overline{q}\,Q\overline{Q}\,+\,n\,g$ (distinct
   flavours) up to NLC accuracy, $\mathcal{O}(N_c^{n})$, in the
   fundamental basis.  The various columns $\min(n_1,n-n_1)$
   correspond to the smallest number of gluons of the two partitions
   $n_1$ and $n-n_1$ as defined in Eq.~\eqref{eq:2qq_amp_types}.
   Within a column, the number before (after) the vertical bar
   correspond to a dual amplitude of type $\A_1$ ($\A_2$), see
   Eq.~\eqref{eq:2qq_amp_types}. The numbers between brackets in the
   final column denote the total number of columns in the colour
   matrix. In bold are marked those numbers which appear twice in the mirrored partition.}
\label{tab:2qq_fund_df}
\end{table}
\endgroup

\begingroup
\renewcommand*{\arraystretch}{1.2}
 \begin{table}[htb!]
 \center
 \begin{tabular}{  l r@{\,,\,}c@{\,\big|\,} lr@{\,,\,}c@{\,\big|\,}l r@{\,,\,}c@{\,\big|\,}l r@{\,,\,}c@{\,\big|\,}l r@{\,,\,}c@{\,\big|\,}l r@{\,,\,}c@{\,\big|\,}l    @{\qquad} l }
 \toprule
 \multicolumn{6}{l}{$q\overline{q}\,Q\overline{Q}\,+\,n\,g$} & \multicolumn{14}{r}{Colour-flow:\qquad $\A_1$, $\A^1_1$ \big| $\A_2$ types} \\
 \multirow{2}{*}{$n$} & \multicolumn{18}{c} {$\min(n_1,n-n_1)$} & \\
& \multicolumn{3}{c}{$0$} & \multicolumn{3}{c}{$1$} & \multicolumn{3}{c}{$2$} & \multicolumn{3}{c}{$3$} & \multicolumn{3}{c}{$4$} & \multicolumn{3}{c}{$5$} &  \\
\midrule
0 & 2& - & 2 & \multicolumn{3}{c}{ } & \multicolumn{3}{c}{ } & \multicolumn{3}{c}{ } & \multicolumn{3}{c}{ } & \multicolumn{3}{c}{ } & (2) \\
1 & \textbf{4} & 1 & \textbf{3} & \multicolumn{3}{c}{ } & \multicolumn{3}{c}{ } & \multicolumn{3}{c}{ } & \multicolumn{3}{c}{ } & \multicolumn{3}{c}{ } & (6) \\
2 & \textbf{9} & 1 & \textbf{4} & 10 & 1 & 5 & \multicolumn{3}{c}{ } & \multicolumn{3}{c}{ } & \multicolumn{3}{c}{ } & \multicolumn{3}{c}{ } & (22) \\
3 & \textbf{20} & 1 & \textbf{5} & \textbf{22} & \textbf{1} & \textbf{7} & \multicolumn{3}{c}{ } & \multicolumn{3}{c}{ } & \multicolumn{3}{c}{ } & \multicolumn{3}{c}{ } & (98) \\
4 & \textbf{41}& 1 & \textbf{6} & \textbf{44} & \textbf{1} & \textbf{9} & 45 & 1 & 10 & \multicolumn{3}{c}{ } & \multicolumn{3}{c}{ } & \multicolumn{3}{c}{ } & (522) \\
5 & \textbf{77} & 1 & \textbf{7} & \textbf{81} & \textbf{1} & \textbf{11} & \textbf{83} & \textbf{1} & \textbf{13} & \multicolumn{3}{c}{ } & \multicolumn{3}{c}{ } & \multicolumn{3}{c}{ } & (3262) \\
6 & \textbf{134} & 1 & \textbf{8} & \textbf{139} & \textbf{1} & \textbf{13} & \textbf{142} & \textbf{1} & \textbf{16}  & 143 & 1 & 17 & \multicolumn{3}{c}{ } & \multicolumn{3}{c}{ } & (23486) \\
7 & \textbf{219} & 1 & \textbf{9} & \textbf{225}  & \textbf{1} & \textbf{15} &\textbf{ 229} & \textbf{1} & \textbf{19} & \textbf{231} & \textbf{1} & \textbf{21} & \multicolumn{3}{c}{ } & \multicolumn{3}{c}{ } & (191802)   \\ 
8 & \textbf{340} & 1 & \textbf{10} & \textbf{347} & \textbf{1} & \textbf{17} &\textbf{352} & \textbf{1} & \textbf{22} & \textbf{355} & \textbf{1} & \textbf{25} & 356 & 1 & 26 & \multicolumn{3}{c}{ } & (1753618)  \\ 
9 & \textbf{506} & 1 & \textbf{11} & \textbf{514} & \textbf{1} & \textbf{19} & \textbf{520} & \textbf{1} & \textbf{25} & \textbf{524} & \textbf{1} & \textbf{29} & \textbf{526} & \textbf{1} & \textbf{31} & \multicolumn{3}{c}{ } & (17755382)   \\ 
10 & \textbf{727} & 1 & \textbf{12} & \textbf{736}  & \textbf{1} & \textbf{21} & \textbf{743} & \textbf{1} & \textbf{28} & \textbf{748} & \textbf{1} & \textbf{33} & \textbf{751} & \textbf{1} & \textbf{36} & 752 & 1 & 37 & (197282022)     \\ 

\bottomrule
\end{tabular}
 \caption{Number of non-zero elements in a single row of the colour
   matrix for $q\overline{q}\,Q\overline{Q}\,+\,n\,g$ (distinct
   flavours) up to NLC accuracy, $\NC^n$,
   in the colour-flow basis. 
    The various columns $\min(n_1,n-n_1)$ correspond to the smallest number of gluons of the two partitions $n_1$ and $n-n_1$ as defined in Eq.~\eqref{eq:2qq_c_and_A}. 
Within a column,
   the numbers correspond to dual amplitudes of types $\A_1$, $\A_1^1$
   and, after the vertical bars, $\A_2$, see
   Eqs.~\eqref{eq:2qq_c_and_A} and \eqref{eq:A11}. The numbers between brackets in
   the final column denote the total number of columns in the colour
   matrix. In bold are marked those numbers which appear twice in the mirrored partition.}
\label{tab:2qq_col_df}
\end{table}
\endgroup

The number of dual conjugate amplitudes that need to be computed for a
single dual amplitude at NLC for the two-quark-pair (distinct-flavour) matrix elements
are listed in Tabs.~\ref{tab:2qq_fund_df}~and~\ref{tab:2qq_col_df} in the fundamental and colour-flow
decompositions, respectively. In the tables, each row corresponds to the number
of gluons in the process, denoted in the first column. The various
columns ($\min(n_1,n-n_1)$) represent the different gluon partitions on the two colour lines, using always the smallest number of gluons on one colour line as label, see also Eq.~\eqref{eq:2qq_amp_types}. Furthermore, the number of
dual conjugate amplitudes that contribute at NLC also depends on the
quark ordering in the amplitude, with the numbers before and after the
vertical bars corresponding to the $\A_1$ and $\A_2$ ordering,
respectively. In the case of the colour-flow decomposition,
Tab.~\ref{tab:2qq_col_df}, this number also depends
on whether the dual amplitude contains an external $\U(1)$ gluon ($\A_1^1$) or
not. The numbers for the cases when the dual amplitude is a $\A_1$ amplitude and $\A_1^1$ amplitude, are separated by a comma in the table. In each of the tables,
the number in the final column in brackets correspond to the total
number of dual conjugate amplitudes that need to be considered in the
full-colour calculation, i.e.~the total number of columns in the
colour matrix. The corresponding tables for the same-flavour case can be obtained in the following way: in the fundamental decomposition, both the number of dual conjugate amplitudes for a row with $\A_1$- and $\A_2$-type diagrams is the same as the number corresponding to a $\A_1$-type dual amplitude  in Tab. ~\ref{tab:2qq_fund_df}, and in the colour-flow decomposition, the number in both the $\A_1$- and $\A_2$-type columns is the same as the number of $\A_1$-type columns in Tab. ~\ref{tab:2qq_col_df}, and similarly both the numbers in the $\A^1_1$- and $\A^1_2$-type columns is the same as the numbers presented in the $\A_1^1$ columns in Tab.~\ref{tab:2qq_col_df}.

Similarly as to the cases of the all-gluon and one-quark-pair matrix
elements, the total number of conjugate dual amplitudes in the
two-quark-pair case also scales factorially with the number of gluons
involved in the matrix elements, when considering the full colour accuracy. The total number is the same in the
distinct-flavour and same-flavour cases, but is considerably worse in
the colour-flow basis as compared to the fundamental basis. As before,
the reason is the need for the extra dual amplitudes with the external
$\U(1)$ gluons in the colour-flow decomposition. Limiting the matrix
elements to NLC accuracy greatly decreases the number of contributing
dual conjugate amplitudes for a single dual amplitude. In fact, for
the worst case, the scaling is still only like a polynomial of degree
$4$. However, this depends on the type of dual amplitude for which
this number is computed. In the distinct-flavour case, the scaling
with the $\A_2$ type is only
linear with the number of gluons $n$; only the $\A_1$ shows the $n^4$
scaling. While there are small differences depending on the partition of gluons on the two colour lines in the colour
ordering, the scaling is independent from this. For the same-flavour
matrix elements there is, of course, no difference for the $\A_1$ and
$\A_2$ dual amplitude types.

\begin{figure}[htb!]
\center
\begin{minipage}{0.48\textwidth}
\center
\includegraphics[scale=0.9]{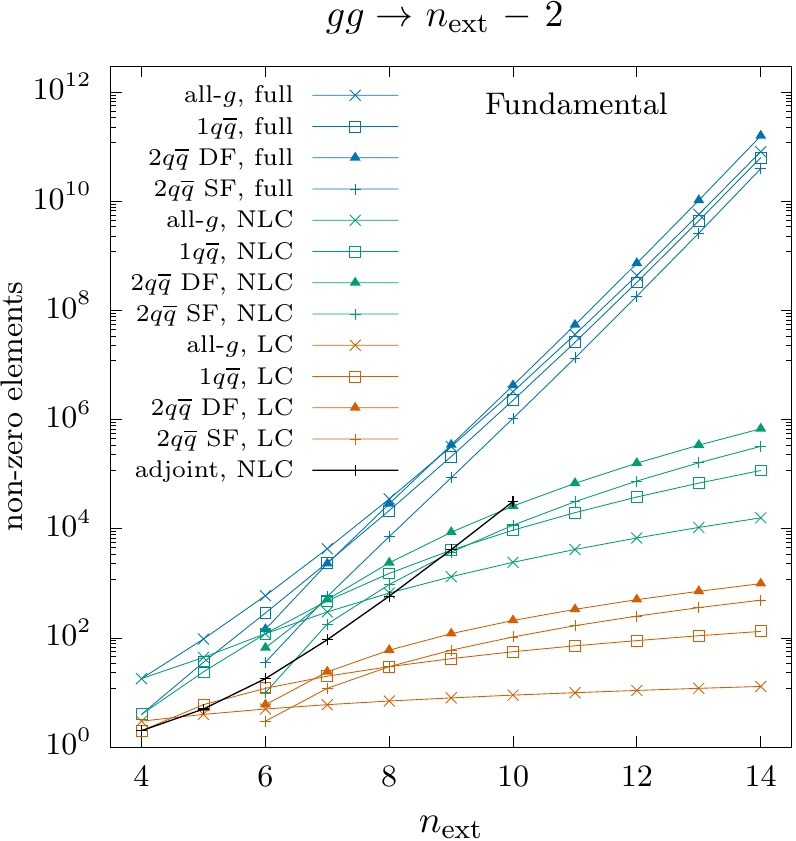}
\end{minipage}
\begin{minipage}{0.48\textwidth}
\center
\includegraphics[scale=0.9]{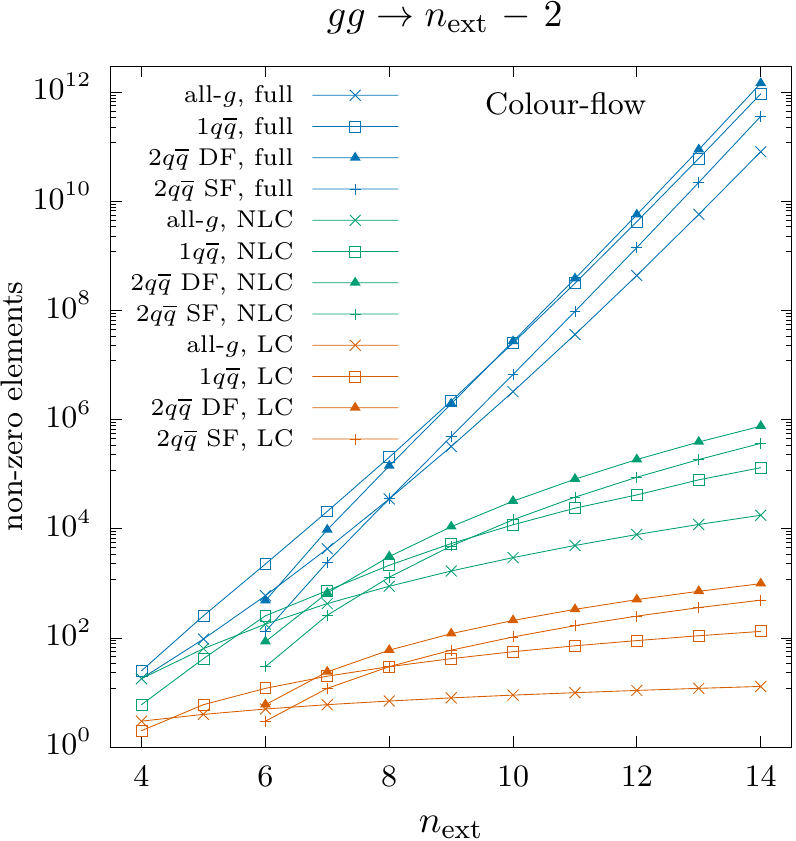}
\end{minipage}
\caption{Number of non-zero elements in the colour matrix at LC (red),
  NLC (green) and full colour (blue) for $gg$ initial state in
  fundamental (left) and colour-flow decomposition (right) for various
  channels: all-gluon (cross), one-quark-pair (box), two-quark-pair
  distinct-flavour (triangle) two-quark-pair same-flavour (bar). In
  the left figure is also shown the number of non-zero elements in the
  adjoint decomposition at NLC accuracy up to 10 external particles. }
\label{fig:gg}
\end{figure}

\begin{figure}[htb!]
\center
\begin{minipage}{0.48\textwidth}
\center
\includegraphics[scale=0.9]{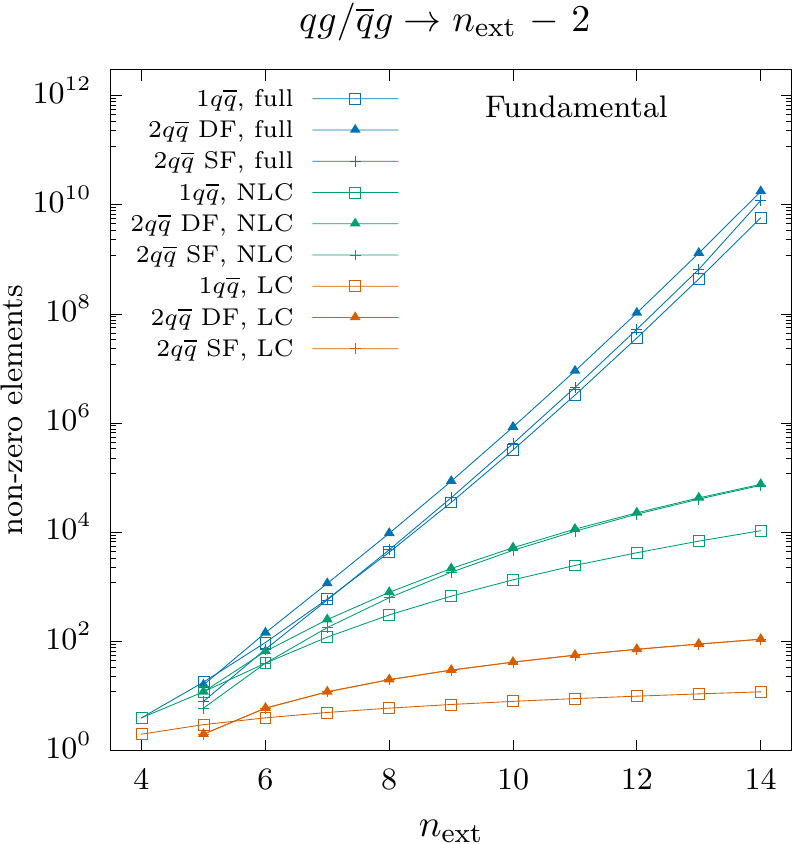}
\end{minipage}
\begin{minipage}{0.48\textwidth}
\center
\includegraphics[scale=0.9]{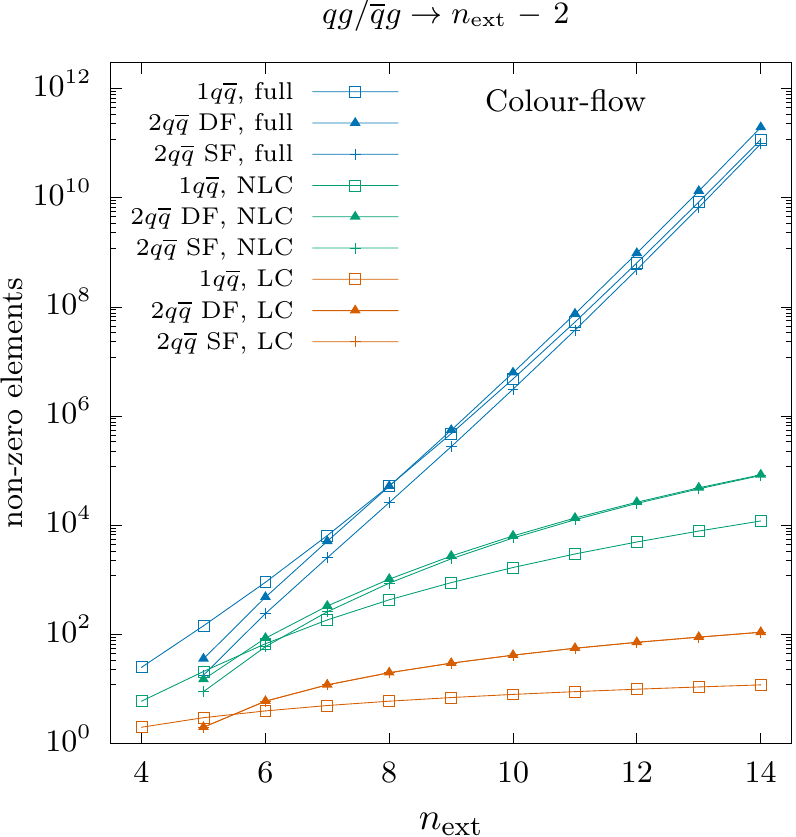}
\end{minipage}
\caption{Number of non-zero elements in the colour matrix at LC (red),
  NLC (green) and full colour (blue) for $qg/\overline{q}g$ initial
  state in fundamental (left) and colour-flow decomposition (right)
  for various channels: one-quark-pair (box), two-quark-pair
  distinct-flavour (triangle) two-quark-pair same-flavour (bar).}
\label{fig:qg}
\end{figure}

\begin{figure}[htb!]
\center
\begin{minipage}{0.48\textwidth}
\center
\includegraphics[scale=0.9]{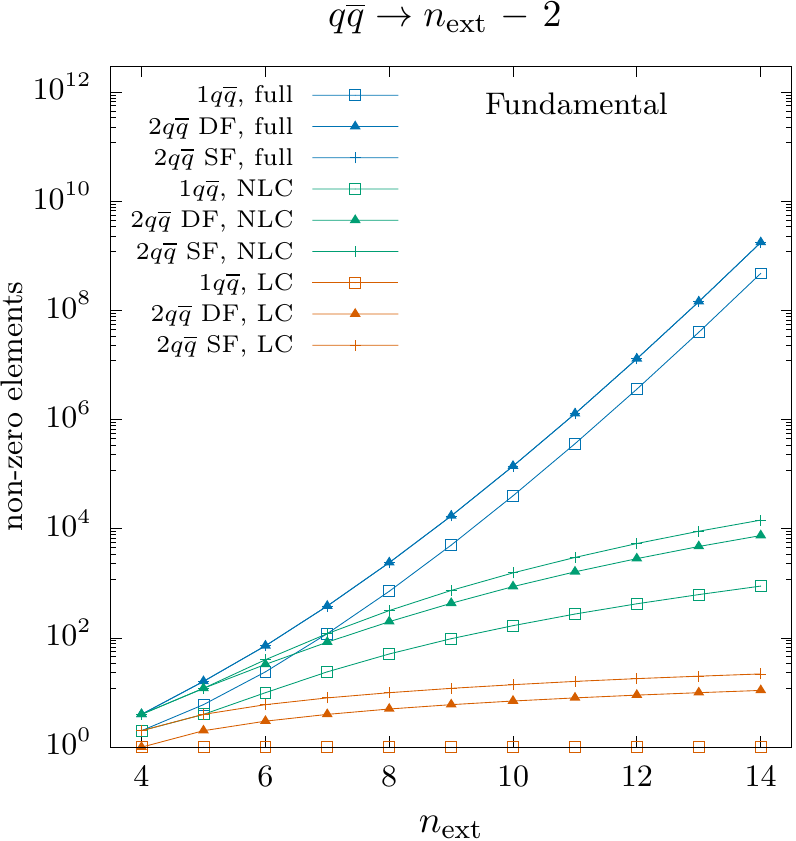}
\end{minipage}
\begin{minipage}{0.48\textwidth}
\center
\includegraphics[scale=0.9]{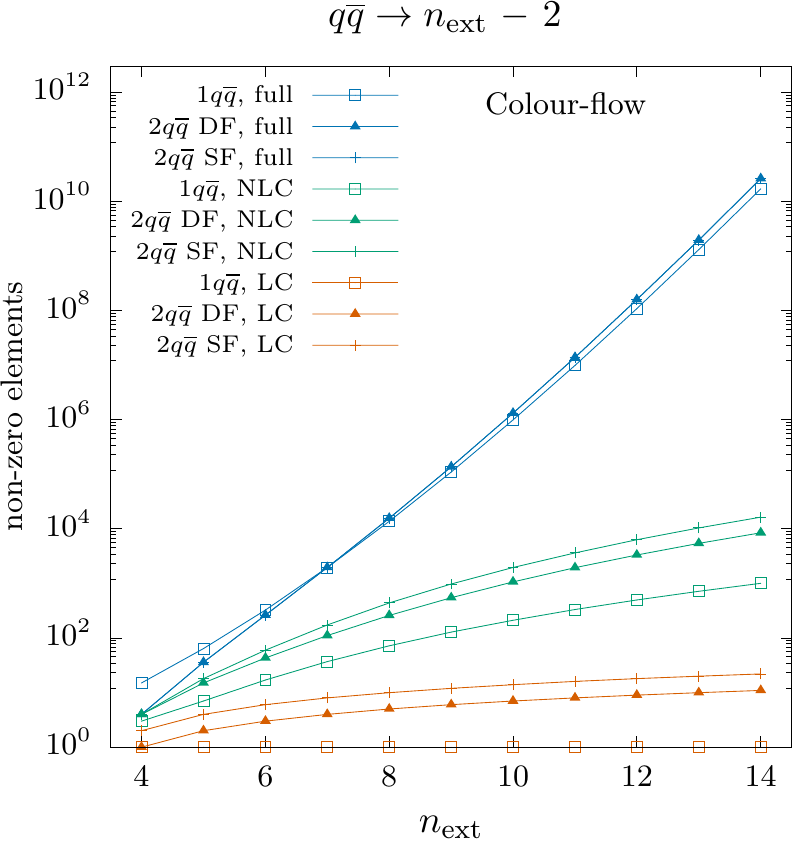}
\end{minipage}
\caption{Number of non-zero elements in the colour matrix at LC (red),
  NLC (green) and full colour (blue) for $q\overline{q}$ initial
  state in fundamental (left) and colour-flow decomposition (right)
  for various channels: one-quark-pair (box), two-quark-pair
  distinct-flavour (triangle) two-quark-pair same-flavour (bar).}
\label{fig:qqbar}
\end{figure}

\begin{figure}[htb!]
\center
\begin{minipage}{0.48\textwidth}
\center
\includegraphics[scale=0.9]{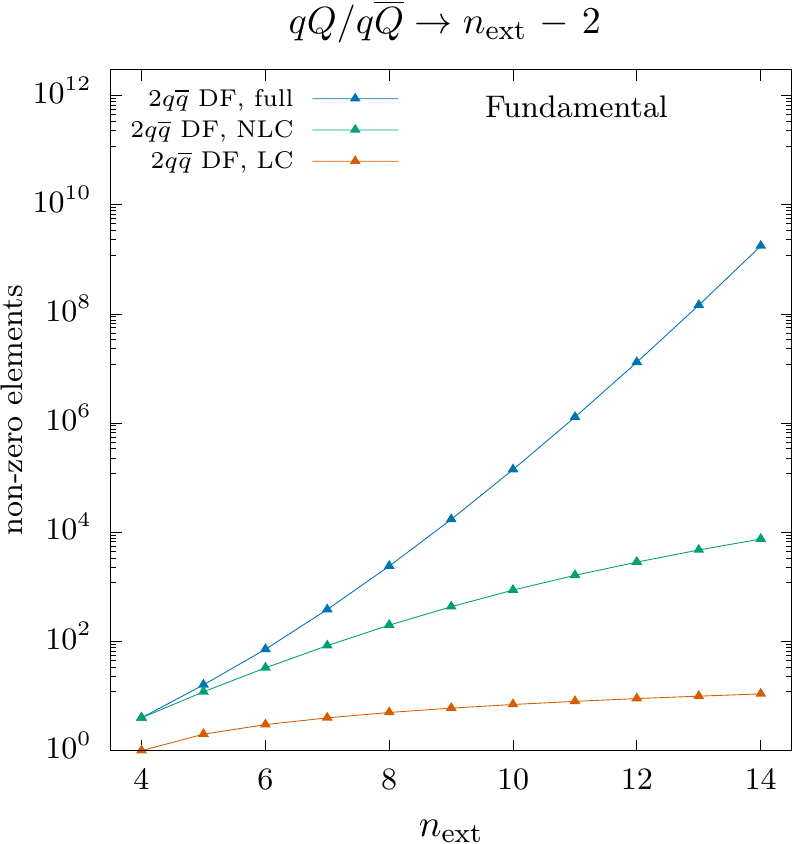}
\end{minipage}
\begin{minipage}{0.48\textwidth}
\center
\includegraphics[scale=0.9]{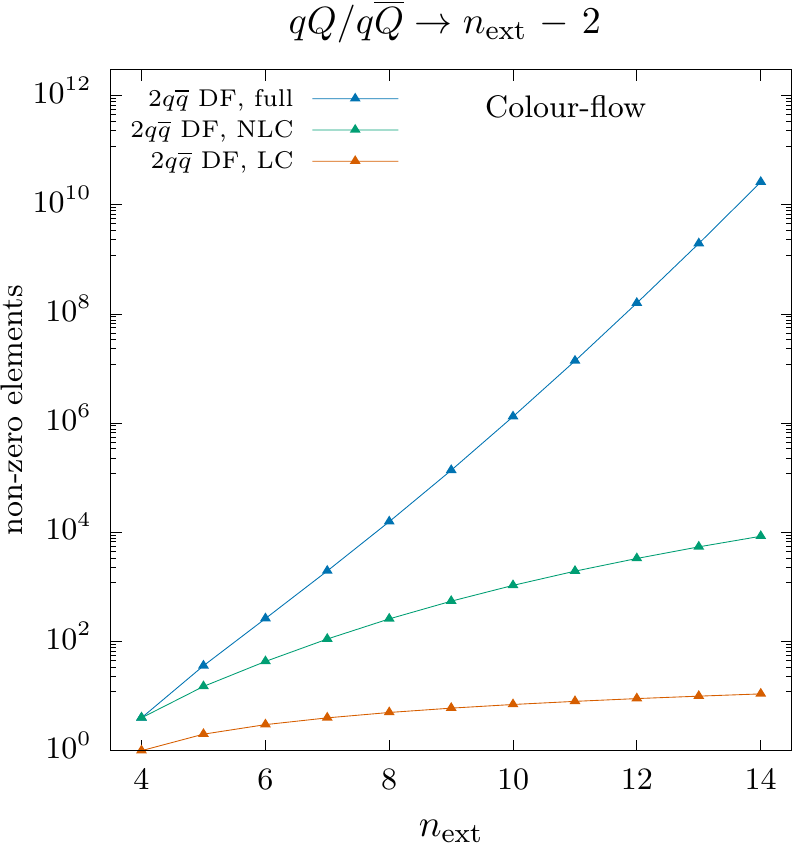}
\end{minipage}
\caption{Number of non-zero elements in the colour matrix at LC (red),
  NLC (green) and full colour (blue) for $qQ/q\overline{Q}$ initial
  state in fundamental (left) and colour-flow decomposition (right)
  for the two-quark-pair distinct-flavour channel (triangle).}
\label{fig:qQ}
\end{figure}

\begin{figure}[htb!]
\center
\begin{minipage}{0.48\textwidth}
\center
\includegraphics[scale=0.9]{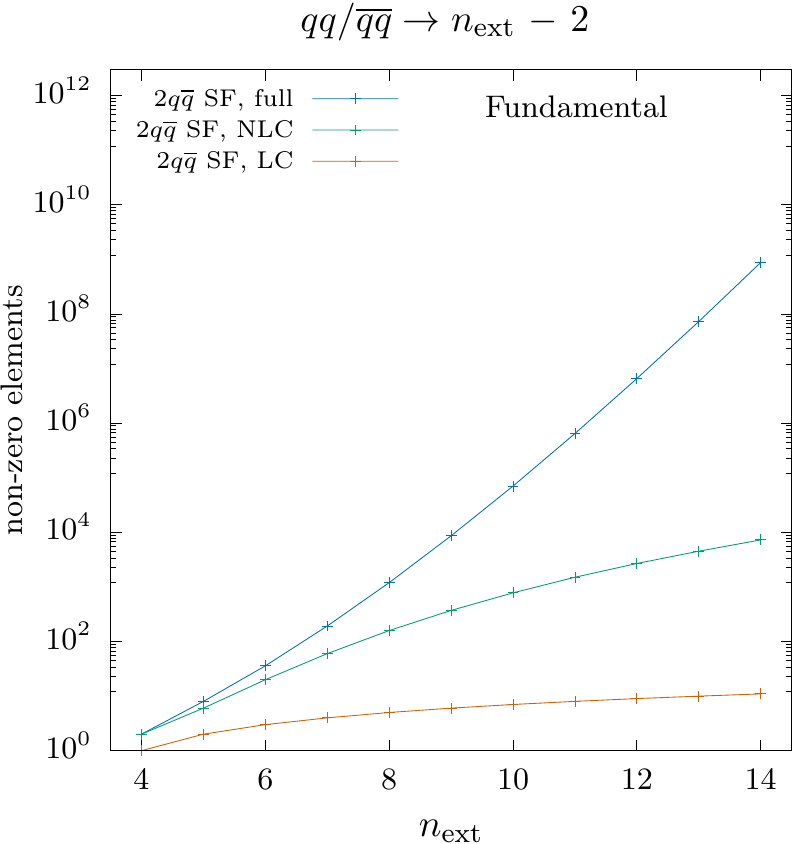}
\end{minipage}
\begin{minipage}{0.48\textwidth}
\center
\includegraphics[scale=0.9]{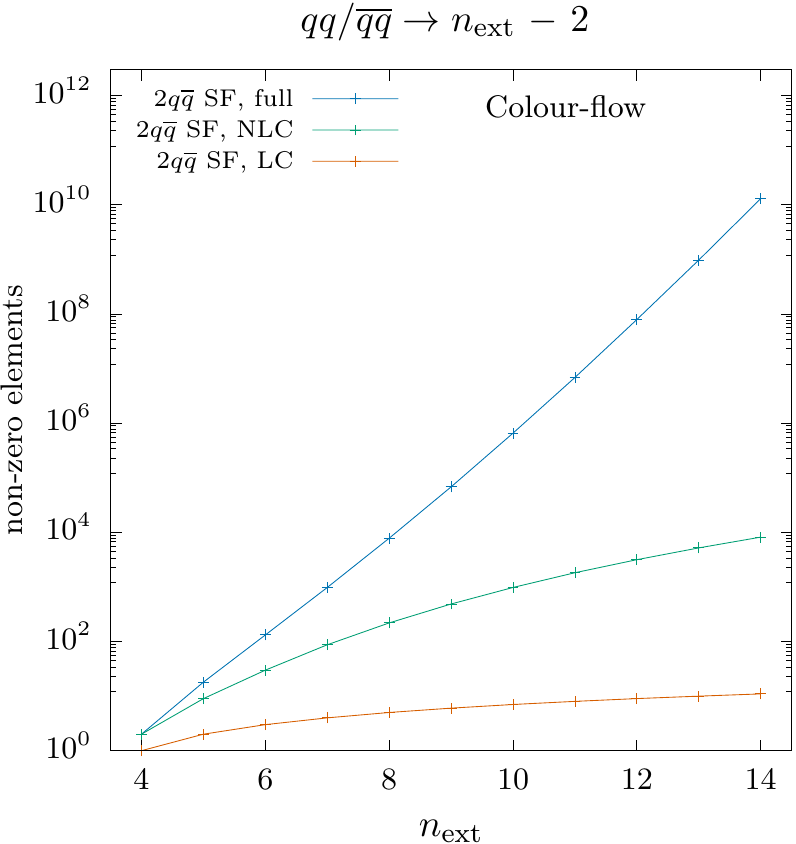}
\end{minipage}
\caption{Number of non-zero elements in the colour matrix at LC (red),
  NLC (green) and full colour (blue) for $qq/\overline{q}\overline{q}$ initial
  state in fundamental (left) and colour-flow decomposition (right)
  for the two-quark-pair same-flavour channel (bar).}
\label{fig:qq}
\end{figure}

As discussed above, in Tabs.~\ref{tab:gluons}-\ref{tab:2qq_col_df} the
number of non-zero terms in a single row of the colour matrix are
listed. The relevant number of rows, including the reduction due to
the symmetrisation of the phase-space, are discussed and given in
Sec.~\ref{sec:symmetry}. Hence, this allows us to determine the main
results of this work: the total number of elements in the double sum
of Eq.~\eqref{eq:colourdecomp} that need to be considered to compute
matrix elements at NLC accuracy. The results are presented in the form
of Figs.~\ref{fig:gg}-\ref{fig:qq} as a function of the number of
external particles, $n_{\rm ext}$. In the figures, the number of
non-zero elements at full-colour accuracy is presented with blue
curves, at NLC accuracy with green curves, and at leading-colour
accuracy with red curves. The curves with crosses correspond to the
all-gluon process, the boxes to the processes with one quark line, the
triangles to processes with two distinct-flavour quark pairs, and the
vertical bars to processes with two same-flavour quark pairs.
Figure~\ref{fig:gg} corresponds to the processes with a $gg$ initial
state. The $qg/\overline{q}g$ initial state processes are plotted in
Fig.~\ref{fig:qg}.  In Fig.~\ref{fig:qqbar} the same curves are given
for $q\overline{q}$ initial state in the two decompositions and the
processes with $qQ/q\overline{Q}$ and $qq/\overline{q}\overline{q}$
initial states are presented in Figs.~\ref{fig:qQ}~and~\ref{fig:qq},
respectively. For all these figures, the left plot corresponds to the
fundamental decomposition and the right plot to the colour-flow
decomposition. In the left figure of Fig.~\ref{fig:gg} we also present
the NLC accurate curve (black) in the adjoint decomposition. For
completeness, the data that fills these figures is given in
Tabs.~\ref{tab:gg_fund}-\ref{tab:qQ_qq} in Appendix~\ref{sec:tables}.

From the figures the following patterns can be deduced.
\begin{itemize}
  \item The factorial growth of the number of terms with the number of
    external particles in the full-colour results is clearly
    visible. This is irrespective of the process considered.
  \item Similarly, it can be seen that all the (N)LC contributions
    scale polynomially with the number of partons participating in the
    scattering process, with the maximum degree at NLC higher than at
    LC. The exception is the NLC in the adjoint representation for the
    all-gluon scattering process, which does not show a polynomial
    scaling.
    \item The turn-over for the efficiency (number of non-zero terms)
      between the fundamental (or colour-flow) and the adjoint
      decomposition for the all-gluon process appears between $n=8$ and
      $n=9$, including the phase-space symmetrisation.
  \item The differences between the fundamental and colour-flow
    decompositions are rather small at (N)LC accuracy, while in the
    full-colour results the colour-flow basis requires the computation
    of significantly more terms. 
  \item At all the accuracies (full-colour, NLC and LC) and for each
    of the initial states, the contributions with the most terms are
    the two-quark-pair distinct-flavour processes. The reason for this is mostly due
    to the greater reduction from the phase-space symmetrisation for
    multi-gluon contributions. The exceptions are the processes
    with the same-flavour $q\overline{q}$ pair initial states, where
    the same-flavour two-quark-line processes require more terms. The
    reason is that additional reduction due to the phase-space
    symmetrisation for the same-flavour two-quark-line matrix elements
    is absent.
  \item In general, already for $2 \to 4,5$ processes there are
    significantly fewer contributions to be considered at NLC than at
    full-colour (note the squared exponential scale for the $y$-axes
    in these plots). This suggests that even at relatively low
    multiplicities the colour expansion up to NLC results in
    improvements in calculation speed compared to
    full-colour. 
\end{itemize}

\section{Conclusion}\label{sec:conclusion}
\noindent
We have presented, at tree-level, the rules for obtaining all dual
conjugate amplitudes for a given dual amplitude which are needed to
perform a NLC accurate computation of scattering probabilities, for
all-gluon, one-quark-pair and two-quark-pair processes. By including
also the phase-space symmetrisation, we have presented the number of
non-zero elements in the NLC accurate colour matrix in both the
fundamental and colour-flow decompositions and found that in both
cases the number of elements in the colour matrix reduces from the
factorial growth (with the number of external particles) at full
colour, to a polynomial growth at NLC. This opens a path to a matrix
element generator of polynomial complexity, without a need for Monte Carlo sampling over colours.

We have compared the fundamental and colour-flow decompositions and
found that the fundamental decomposition is slightly more efficient
for making the colour matrix sparse at NLC, which is partly due to
additional dual amplitudes with external $\U(1)$ gluons which are
present in the colour-flow basis only, and partly due to the fewer
exceptions to the general rules in the colour-flow basis. We also
compare the sparseness of the colour matrix in these non-minimal bases
with the adjoint decomposition for the all-gluon case and find
that at $n=9$ gluons or more, the fundamental and colour-flow
decompositions are more efficient than the adjoint basis at NLC.

Already for scattering amplitudes at moderate multiplicities, $2\to
4,5$, there are significantly fewer contributions at NLC as compared
to full colour. This suggests that this work could also be of
practical interest for computing cross sections and differential
distributions for LHC scattering processes. It is worth investigating
the exact trade-off between the increase in speed and the decrease in
accuracy due to omission of contributions beyond NLC.  We formally suppress orders of $1/\NC^3$ (corresponding to a few percent; although this might be enhanced for high-multiplicity processes), therefore we expect other sources of uncertainties, such as renormalization scale dependence, to be dominant. An in-depth analysis of this is left for future work.

All the rules for NLC are presented for processes with up two two quark pairs, at tree level only. While the set of rules can be extended to cover also processes with larger number of quark pairs, these processes are phenomenologically less relevant. The loop-level colour structures have been worked out in great detail and available in literature 
for the all-gluon and multiple-quark-pair cases, however they
might pose further levels of complexity for determining the NLC elements
in the colour matrix.

The code for the colour computations presented in the paper is
available from the authors upon request.

\section*{Acknowledgements}
The authors thank Stefano Frixione and TV thanks Andrew Lifson for useful discussions.
 This work is supported by the Swedish Research Council under contract
number 2016-05996.

\newpage
\appendix

\section{Data for figures}\label{sec:tables}
In this appendix we list the numbers in
Tabs.~\ref{tab:gg_fund}-\ref{tab:qQ_qq} that were used to create  Figs.~\ref{fig:gg}-\ref{fig:qq}.

\begingroup
\renewcommand*{\arraystretch}{1.2}
 \begin{table}[htb!]
 \center
 \begin{tabular}{  r   r @{~} l  r @{~} l  r @{~} l  r @{~} l  }
\toprule
 \multicolumn{8}{l} {$gg \rightarrow n_{\rm ext}-2$  \quad \quad \quad \quad \quad \quad \quad \quad Fundamental} \\
 $n_{\rm ext}$ & \multicolumn{2}{c}{all-gluon} & \multicolumn{2}{c}{one-quark} &  \multicolumn{2}{c}{two-quark DF} & \multicolumn{2}{c}{two-quark SF}   \\
 \midrule
 4  &    18 &  (3) &      6 &   (2) &      - &       &      - &       \\ 
 5  &    44 &  (4) &     24 &   (6) &      - &       &      - &       \\ 
 6  &   120 &  (5) &    120 &  (12) &     66 &   (6) &     10 &   (3) \\ 
 7  &   300 &  (6) &    480 &  (20) &    504 &  (24) &    180 &  (12) \\ 
 8  &   665 &  (7) &   1530 &  (30) &   2388 &  (60) &    954 &  (30) \\ 
 9  &  1328 &  (8) &   4074 &  (42) &   8680 & (120) &   3720 &  (60) \\
 10 &  2439 &  (9) &   9464 &  (56) &  26160 & (210) &  11715 & (105) \\
 11 &  4190 & (10) &  19800 &  (72) &  68376 & (336) &  31500 & (168) \\
 12 &  6820 & (11) &  38160 &  (90) & 159824 & (504) &  75040 & (252) \\
 13 & 10620 & (12) &  68860 & (110) & 341568 & (720) & 162504 & (360) \\
 14 & 15938 & (13) & 117744 & (132) & 678510 & (990) & 325890 & (495) \\
 \bottomrule
\end{tabular}
 \caption{Number of non-zero elements in the colour matrix at NLC and
   at LC (in parenthesis) including the reduction due to the
   phase-space symmetrisation, for $gg$ initial states for all-gluon,
   one-quark-pair and two-quark-pair (same-flavour (SF) and
   distinct-flavour (DF)) processes for various numbers of final-state
   particles ($n_{\rm ext}-2$) in the fundamental representation.}
\label{tab:gg_fund}
\end{table}
\endgroup

\begingroup
\renewcommand*{\arraystretch}{1.2}
 \begin{table}[htb!]
 \center
 \begin{tabular}{  r   r @{~} l   r @{~} l   r @{~} l   r @{~} l  }
\toprule
 \multicolumn{8}{l} {$gg \rightarrow n_{\rm ext}-2$  \quad \quad \quad \quad \quad \quad \quad \quad Colour-flow} \\
 $n_{\rm ext}$ & \multicolumn{2}{c}{all-gluon} & \multicolumn{2}{c}{one-quark} &  \multicolumn{2}{c}{two-quark DF} & \multicolumn{2}{c}{two-quark SF}   \\
 \midrule
 4  &    18 &  (3) &     6 &   (2) &      - &       &      - &       \\ 
 5  &    64 &  (4) &     42 &   (6) &      - &       &      - &       \\ 
 6  &   180 &  (5) &    255 &  (12) &    86 &   (6) &     30 &   (3) \\ 
 7  &   426 &  (6) &    740 &  (20) &    666 &  (24) &    261 &  (12) \\ 
 8  &   889 &  (7) &   2160 &  (30) &   3108 &  (60) &   1302 &  (30) \\ 
 9  &  1688 &  (8) &   5376 &  (42) &  10980 & (120) &   4870 &  (60) \\
 10 &  2979 &  (9) &  11872 &  (56) &  32100 & (210) &  14685 & (105) \\
 11 &  4960 & (10) &  23904 &  (72) &  81606 & (336) &  38124 & (168) \\
 12 &  7876 & (11) &  44730 &  (90) & 186256 & (504) &  88256 & (252) \\
 13 & 12024 & (12) &  78870 & (110) & 390168 & (720) & 186884 & (360) \\
 14 & 17758 & (13) & 132396 & (132) & 762210 & (990) & 367740 & (495) \\
 \bottomrule
\end{tabular}
 \caption{Number of non-zero elements in the colour matrix at NLC and
   at LC (in parenthesis) including the reduction due to the
   phase-space symmetrisation, for $gg$ initial states for all-gluon,
   one-quark-pair and two-quark-pair (same-flavour (SF) and
   distinct-flavour (DF)) processes for various numbers of final-state
   particles ($n_{\rm ext}-2$) in the colour-flow representation.}
\label{tab:gg_col}
\end{table}
\endgroup

\begingroup
\renewcommand*{\arraystretch}{1.2}
 \begin{table}[htb!]
 \center
 \begin{tabular}{  r   r @{~} l r @{~} l  r @{~} l   r @{~} l r @{~} l r @{~} l}
\toprule
\multicolumn{7}{l}{$qg/\overline{q}g  \rightarrow n_{\rm ext}-2$  \quad \quad \quad Fundamental} &  \multicolumn{6}{c} { Colour-flow} \\
 $n_{\rm ext}$ & \multicolumn{2}{c}{one-quark} &  \multicolumn{2}{c}{two-quark DF}&  \multicolumn{2}{c}{two-quark SF}  &  \multicolumn{2}{c}{one-quark} &    \multicolumn{2}{c}{two-quark DF} &  \multicolumn{2}{c}{two-quark SF}   \\
 \midrule
  4 &     4 &  (2) &     - &       &     - &       &    6 &  (2) &     - &       &     - &      \\ 
  5 &    12 &  (3) &    12 &   (2) &     6 &   (2) &    21 &  (3) &    15 &   (2) &    9 &   (2)  \\
  6 &    40 &  (4) &    66 &   (6) &    40 &   (6) &   68 &  (4) &   86 &   (6) &    60 &   (6)  \\ 
  7 &   120 &  (5) &   252 &  (12) &   180 &  (12) &   185 &  (5) &   333 &  (12) &   261 &  (12)   \\ 
  8 &   306 &  (6) &   796 &  (20) &   636 &  (20) &   432 &  (6) &  1036 &  (20) &  876 &  (20)  \\ 
  9 &   679 &  (7) &  2170 &  (30) &  1860 &  (30) &   896 &  (7) &  2745 &  (30) &  2435 &  (30)    \\
 10 &  1352 &  (8) &  5232 &  (42) &  4686 &  (42) &  1696 &  (8) &  6420 &  (42) &  5874 &  (42)  \\
 11 &  2475 &  (9) & 11396 &  (56) & 10500 &  (56) &  2988 &  (9) & 13601 &  (56) & 12705 &  (56)  \\
 12 &  4240 & (10) & 22832 &  (72) & 21440 &  (72) &  4970 & (10) & 26608 &  (72) & 25216 &  (72)   \\
 13 &  6886 & (11) & 42696 &  (90) & 40626 &  (90) &  7887 & (11) & 48771 &  (90) & 46701 &  (90)   \\
 14 & 10704 & (12) & 75390 & (110) & 72420 & (110) & 12036 & (12) & 84690 & (110) & 81720 & (110) \\
 \bottomrule
\end{tabular}
 \caption{Number of non-zero elements in the colour matrix at NLC and
   at LC (in parenthesis) including the reduction due to the
   phase-space symmetrisation, for $qg/\overline{q}g$ initial states
   for one-quark-pair and two-quark-pair (same-flavour (SF) and
   distinct-flavour (DF)) processes for various numbers of final-state
   particles ($n_{\rm ext}-2$) in the fundamental and colour-flow
   representations.}
\label{tab:qg}
\end{table}
\endgroup

\begingroup
\renewcommand*{\arraystretch}{1.2}
 \begin{table}[htb!]
 \center
 \begin{tabular}{  r   r @{~} l r @{~} l  r @{~} l   r @{~} l r @{~} l r @{~} l}
\toprule
\multicolumn{7}{l}{$q\overline{q}  \rightarrow n_{\rm ext}-2$  \quad \quad \quad Fundamental} &  \multicolumn{6}{c} { Colour-flow} \\
 $n_{\rm ext}$ & \multicolumn{2}{c}{one-quark} &  \multicolumn{2}{c}{two-quark DF}&  \multicolumn{2}{c}{two-quark SF}  &  \multicolumn{2}{c}{one-quark} &    \multicolumn{2}{c}{two-quark DF} &  \multicolumn{2}{c}{two-quark SF} \\
 \midrule
 4 &  2- & (1)  & 4 & (1)& 4  & (2)  &  3 & (1)  & 4 & (1)& 4 & (2)    \\ 
 5 &  4 & (1)  & 12 & (2) & 12 & (4)  &  7 & (1)  & 15  & (2) & 18 & (4) \\ 
 6 &  10 & (1)   & 33 & (3)&   40 & (6) &  17 & (1)   & 43 & (3)&   60  & (6)  \\ 
 7 & 24 & (1)  & 84 & (4) & 120 & (8) & 37 & (1)  & 111 & (4) &  174 & (8)   \\ 
 8 &  51 & (1)   &  199  & (5) & 318 & (10)    &  72 &  (1)   &  259 &  (5) &  438  & (10)    \\ 
 9 & 97 & (1)    & 434 & (6)  & 744 & (12)  & 128 & (1)    & 549 & (6)  &  974 & (12)   \\
 10 &  169 & (1)   & 872 & (7)  & 1562 & (14) &  212 & (1)   & 1070 & (7)  & 1958 & (14)   \\
 11 &  275 & (1)   & 1628 & (8)  & 3000 & (16)  &  332 & (1)   & 1943 & (8)  & 3630 & (16) \\
 12 &  424 & (1)   & 2854 & (9)  & 5360 & (18) &  497 & (1)   & 3326 & (9)  & 6304 & (18)  \\
 13 &  626 & (1)   & 4744 & (10)  & 9028 & (20) &  717 & (1)   & 5419 & (10)  & 10378 & (20)   \\
 14 &  892 & (1)   & 7539 & (11)  & 14484 & (22) &  1003 & (1)   & 8469 & (11)  & 16344 & (22)   \\
 \bottomrule
\end{tabular}
 \caption{Number of non-zero elements in the colour matrix at NLC and
   at LC (in parenthesis) including the reduction due to the
   phase-space symmetrisation, for $q\overline{q}$ initial state
   for one-quark-pair and two-quark-pair (same-flavour (SF) and
   distinct-flavour (DF)) processes for various numbers of final-state
   particles ($n_{\rm ext}-2$) in the fundamental and colour-flow
   representations.}
\label{tab:qqbar}
\end{table}
\endgroup

\begingroup
\renewcommand*{\arraystretch}{1.2}
 \begin{table}[htb!]
 \centering
     \begin{minipage}{.48\linewidth}
     \center
 \begin{tabular}{   r   r @{~} l r @{~} l  }
 \toprule
 \multicolumn{3}{l} {$qQ/q\overline{Q}$ \quad \quad Fundamental} & \multicolumn{2}{c} {Colour-flow}\\
 $n_{\rm ext}$ & \multicolumn{2}{c}{two-quark DF} & \multicolumn{2}{c}{two-quark DF} \\
 \midrule
  4 &    4 &  (1) &    4 &  (1) \\
  5 &   12 &  (2) &   15 &  (2) \\
  6 &   33 &  (3) &   43 &  (3) \\
  7 &   84 &  (4) &  111 &  (4) \\
  8 &  199 &  (5) &  259 &  (5) \\
  9 &  434 &  (6) &  549 &  (6) \\
 10 &  872 &  (7) & 1070 &  (7) \\
 11 & 1628 &  (8) & 1943 &  (8) \\
 12 & 2854 &  (9) & 3326 &  (9) \\
 13 & 4744 & (10) & 5419 & (10) \\
 14 & 7539 & (11) & 8469 & (11) \\
 \bottomrule
\end{tabular}
\end{minipage}
\begin{minipage}{.48\linewidth}
 \begin{tabular}{   r   r @{~} l r @{~} l  }
\toprule \multicolumn{3}{l} {$qq/\overline{q}\overline{q}$ \quad \quad
  Fundamental} & \multicolumn{2}{c} {Colour-flow}\\ $n_{\rm ext}$ &
\multicolumn{2}{c}{two-quark SF} & \multicolumn{2}{c}{two-quark SF} \\
\midrule
 4  &    2 &  (1) &    2 &  (1) \\
 5  &    6 &  (2) &   9 &  (2) \\
 6  &   20 &  (3) &   30 &  (3) \\
 7  &   60 &  (4) &  87 &  (4) \\
 8  &  159 &  (5) &  219 &  (5) \\
 9  &  372 &  (6) &  487 &  (6) \\
 10 &  781 &  (7) & 979 &  (7) \\
 11 & 1500 &  (8) & 1815 &  (8) \\
 12 & 2680 &  (9) & 3152 &  (9) \\
 13 & 4514 & (10) & 5189 & (10) \\
 14 & 7242 & (11) & 8172 & (11) \\
\bottomrule
\end{tabular}
\end{minipage}
 \caption{Number of non-zero elements in the colour matrix at NLC and
   at LC (in parenthesis) including the reduction due to the
   phase-space symmetrisation, for $qQ/q\overline{Q}$ (left) and
   $qq/\overline{q}\overline{q}$ (right) initial states for
   two-quark-pair processes for various numbers of final-state
   particles ($n_{\rm ext}-2$) in the fundamental and colour-flow
   representations.}
\label{tab:qQ_qq}
\end{table}
\endgroup

\newpage
\bibliography{Color_30}{}
\bibliographystyle{spphys}

\end{document}